 \documentclass[proof]{pasj01}
\Received{$\langle$reception date$\rangle$}
\Accepted{$\langle$acception date$\rangle$}
\Published{$\langle$publication date$\rangle$}
\usepackage{lineno}
\usepackage[figuresright]{rotating}

\begin{document}

\title{Sulphur isotopes toward Sagittarius B2 extended envelope in the Galactic Center}

\author{Qingxu \textsc{Li} \altaffilmark{1}}
\author{Juan \textsc{Li} \altaffilmark{2,3} }
\email{lijuan@shao.ac.cn}
\author{ Siqi  \textsc{Zheng} \altaffilmark{2,3,4}   }
\author{  Junzhi  \textsc{Wang} \altaffilmark {5}}
\author{  Feng  \textsc{Gao} \altaffilmark {6,7} }
 \author{ Yajun  \textsc{Wu} \altaffilmark {2,3}  }

\altaffiltext{1}{400 Baihua Street, Xuhui District, Shanghai 200030, PR China}  
\altaffiltext{2}{Department of Radio Science and Technology, Shanghai Astronomical observatory, 80 Nandan RD, Shanghai 200030, China}
\altaffiltext{3}{Key Laboratory of Radio Astronomy and Technology, Chinese Academy of Sciences, A20 Datun Road, Chaoyang District, Beijing, 100101, P. R. China}
\altaffiltext{4}{University of Chinese Academy of Sciences, 19A Yuquanlu, Beijing 100049, PR China}
\altaffiltext{5}{School of Physical Science and Technology, Guangxi University, Nanning 530004, PR China}
\altaffiltext{6}{Hamburger Sternwarte, Universität Hamburg, Gojenbergsweg 112, 21029, Hamburg, Germany }
\altaffiltext{7}{Max Planck Institute for Extraterrestrial Physics, Giessenbachstraße 1, 85748 Garching, Germany}

\KeyWords{ISM: abundances ${}_1$ ---  ISM: clouds${}_2$ ---SM: individual (Saggitarius B2)${}_3$ --- ISM: 
molecules${}_4$ --- radio lines: ISM${}_5$  }

\maketitle

\begin{abstract}
The isotopic ratios are good tools for probing the stellar nucleosynthesis and chemical evolution. We performed high-sensitivity mapping observations of the J=7-6 rotational transitions of OCS, OC$^{34}$S,  O$^{13}$CS, and OC$^{33}$S toward the Galactic Center giant molecular cloud, Sagittarius B2 (Sgr B2) with IRAM 30m telescope. Positions with optically thin and uncontaminated lines are chosen to determine the sulfur isotope ratios. A $^{32}S/^{34}S$ ratio of 17.1$\pm$0.9 was derived with OCS and OC$^{34}$S lines, while $^{34}S/^{33}S$ ratio of 6.8$\pm$1.9 was derived directly from integrated intensity ratio of OC$^{34}$S and OC$^{33}$S. With independent and accurate measurements of $^{32}S/^{34}S$ ratio, our results confirm the termination of the decreasing trend of $^{32}S/^{34}S$ ratios toward the Galactic Center, suggesting a drop in the production of massive stars at the Galactic centre.
\end{abstract}

\section{Introduction}

The isotopic ratios are good tools for probing the stellar nucleosynthesis and chemical evolution of the Universe. The isotopic abundance ratios of C, N and O could help understand the stellar evolution and nucleosynthesis process \citep{wilson94, maiolino2019, romano22, ou23}. As the main supplementary tool of C, N, and O isotopic abundance ratios, sulphur isotopes could provide information on the late evolutionary stages of massive stars and supernovae (SNe) of Type Ib/c and II, such as oxygen-burning, neon-burning, and s-process nucleosynthesis \citep{wilson94, chin96, humire20}. There are four stable isotopes of S, $^{32}$S, $^{34}$S, $^{33}$S and $^{36}$S. $^{32}$S and $^{34}$S are synthesized by oxygen-burning, while $^{33}$S is synthesized in explosive oxygen- and neon-burning, which is also related to massive stars \citep{chin96,yan23}. 

Among sulphur-bearing molecules, the abundant CS and its isotopologues are widely used to measure sulfur isotope ratios as it is strong and widespread in space \citep{chin96, humire20, yu20,yan23}. A relation between $^{32}$S/$^{34}$S and their galactocentric distance ($D_{GC}$) of $^{32}$S/$^{34}$S=($3.3\pm 0.5$)($D_{GC}$/kpc) + ($4.1\pm3.1$) was found by using a linear least-squares fit to the unweighted data of CS and its isotopologues, while no correlation was obtained between $^{33}$S/$^{34}$S ratios and $D_{GC}$ \citep{chin96, yu20}. Recently, \citet{yan23} confirmed $^{32}$S/$^{34}$S gradients as a function of galactocentric distance with rare isotopologues of CS. 

Such ratios in the Galactic Center regions are specially important, since they can trace star formation history  there. Inconsistent results were reported in the literature \citep{humire20,yu20,yan23}. \citet{humire20} measured the carbon and sulphur abundances toward the +50 km s$^{-1}$ Cloud and several line-of-sight (here after  l.o.s) clouds towards Sgr B2(N) with CS isotopologues. They obtained  $^{32}$S/$^{34}$S ratio of 16.3$^{+2.1}_{-1.7}$ for the +50 km s$^{-1}$ Cloud and  17.9$\pm$5.0 for Sgr B2(N), which indicates a termination of the decreasing tendency at least at a galactocentric distance of 130 pc. However, \citet{yu20} reported a $^{32}$S/$^{34}$S ratio of 7.1 $\pm$ 4.1 with CS isotopologues toward Sgr B2(N), while \citet{yan23} obtained three independent values for three different velocity components of CS isotopologues toward Sgr B2(N), with $^{32}$S/$^{34}$S ratios ranging from 10 to 26.
 
As a common sulfur-bearing molecule in dense clouds, carbonyl sulfide (OCS) and its isotopologues can be an independent probe to measure sulfur isotope ratios. \citet{goldsmith81} once reported a $^{32}$S/$^{34}$S ratio of $\sim$16 for Sgr B2(N) from the OCS/OC$^{34}$S ratio, while \citet{armijos15} reported $^{32}$S/$^{34}$S ratios of $\leq$22 and 8.7$\pm$1.3 in l.o.s clouds towards Sgr A and G+0.693  with OCS/OC$^{34}$S ratio. Such results showed significant difference to each other and to that measured by CS and its isotopologues \citep{humire20,yu20,yan23}. On the other hand, no measurement of $^{33}$S/$^{34}$S ratio with OCS lines was reported in the literature.

Sgr B2 is a giant molecular cloud located in the Galactic Center. It contains two main sites of star-formation, Sgr B2(N) and Sgr B2(M) \citep{belloche13}. There are abundant hot cores, HII regions as well as cold extended envelope in Sgr B2(N) and Sgr B2(M) \citep{bonfand17}. It is noted that previous measurements of sulphur isotope ratios all pointed toward Sgr B2(N) \citep{humire20, yu20, yan23}. However, the strong continuum emission and optical depth of molecular lines make it hard to accurately determine the isotopic ratios. As such, observations with accurate measurement  are required to better obtain value of $^{32}$S/$^{34}$S ratio in the Galactic Center region than that in the literature, which is important to constrain models of stellar interiors and models of the chemical evolution of the Galaxy \citep{kobayashi11}. In this paper, we present mapping observations J=7-6 transitions of OCS, OC$^{34}$S, O$^{13}$CS, and OC$^{33}$S toward Sgr B2 extended envelope. We choose positions in which OCS 7-6 transitions were optically thin to obtain the isotropic ratios of sulphur atoms. The outline of this article is presented as followed. In Section 2, we describe the observations and data reduction. In Section 3, we give the mapping results of OCS, O$^{13}$CS, OC$^{33}$S, OC$^{34}$S, and measured isotopic ratios from OCS species. Scientific discussions are presented  in Section 4, while the conclusion of this paper is given in Section 5.

\section{OBSERVATIONS AND DATA REDUCTION}\label{sec:obs}

We performed point-by-point spectroscopic mapping observations towards Sgr B2 in 2019 May with the IRAM 30m telescope on Pica Veleta, Spain (project 170-18, PI: Feng Gao). The observations was performed at 3-mm band. Position-switching mode was used. The broad-band Eight MIxer Receiver and the FFTSs in FTS200 mode were adopted for the observation. The covered frequency range is 82.3-90 GHz. The channel spacing is 0.195 MHz, corresponding to a velocity resolution of 0.641 km s$^{-1}$ at 84 GHz. The velocity resolution was smoothed to $\sim$ 1.4 km s$^{-1}$ to improve the signal-to-noise ratio for most spectra, while it was smoothed to 2.8 km s$^{-1}$ for O$^{13}$CS toward (60, -30). The telescope pointing was checked every $\sim$ 2 hours on 1757-240. The telescope focus was optimized on 1757-240 at the beginning of the observation.
The integration times range from 24 minutes to 98 minutes for different positions, with typical system temperatures of $\sim$ 110K,  leading to 1$\sigma$ rms in T$_A^*$ of 4-8 mK derived with  the line free channels.  

The observing center is Sgr B2(N) ($\alpha_{J2000}=17^h47^m20^s.0$,$\delta_{J2000}=-28^{\circ}22\arcmin 19.0\arcsec$), with a sampling interval of 30\arcsec. The off position was ($\delta \alpha$, $\delta \beta$)=(-752\arcsec, 342\arcsec) away from Sgr B2(N) \citep{belloche13}. 63 positions was observed in total. The observed transitions are listed in Table \ref{table1}, including J=7-6 transitions of OCS, O$^{13}$CS, OC$^{34}$S, and OC$^{33}$S. The spectroscopic parameters of molecules are taken from CDMS catalog \citep{2005JMoSt.742..215M}. The data processing was conducted using \textbf{GILDAS} software package\footnote{\tt http://www.iram.fr/IRAMFR/GILDAS.}, including CLASS and GREG. The linear baseline subtractions were used for all the spectra. The line parameters are obtained by Gaussian fitting. The data are presented in the unit of antenna temperature (T$^*_A$).

\section{RESULTS}\label{sec:results}

\subsection{Target Selection for deriving  $^{32}$S/$^{34}$S ratio}

The rest frequencies for OCS and its isotopologues have been listed in Table \ref{table1}. The FWHM of lines toward most positions are around 20 km s$^{-1}$. According to 3 mm line survey results of Sgr B2 \citep{belloche13}, OCS and OC$^{33}$S line are clean, with nearby transitions more than 20 km s$^{-1}$ away. For O$^{13}$CS lines, there is a CH$_3$COCH$_3$ 20(10,10)-20(9,11)AE line (84737.075 MHz) $\sim$5.5 km s$^{-1}$ away. However, the E$_l$ of CH$_3$COCH$_3$ is 165 K. Thus it only affects lines toward hot cores, including Sgr B2(N) and Sgr B2(M). We also found that a NH$_2$D 8(3, 6)0s- 8(2, 6)0a line at 83060.244 MHz, which is only 2.3 MHz ($\sim$8.3 km s$^{-1}$) away from OC$^{34}$S, could contaminate OC$^{34}$S lines. 

\begin{figure*}
\centering
\includegraphics[width=0.95\textwidth,angle=-90]{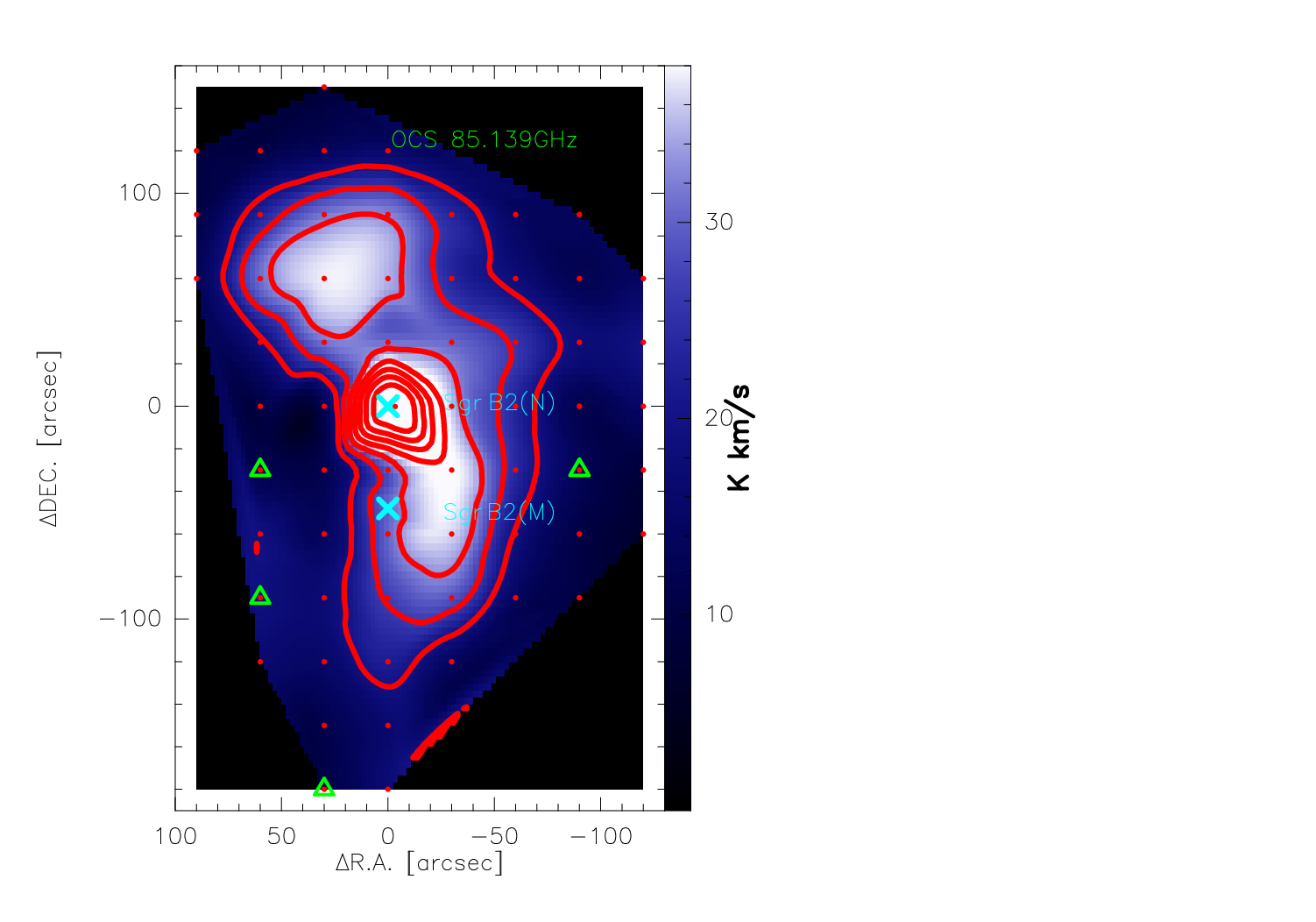}
\caption{The integrated intensity of OCS 7-6 around Sgr B2. The contours represent 30\% to 90\% of integrated intensities, which is 65 K km s$^{-1}$. The red dots denote sampling positions. The crosses denote position of Sgr B2(N) and Sgr B2(M). The green triangles denote selected positions used for calculation of sulphur isotope ratios.}
\label{map}
\end{figure*}

\begin{figure*}
\centering
\includegraphics[width=0.45\textwidth]{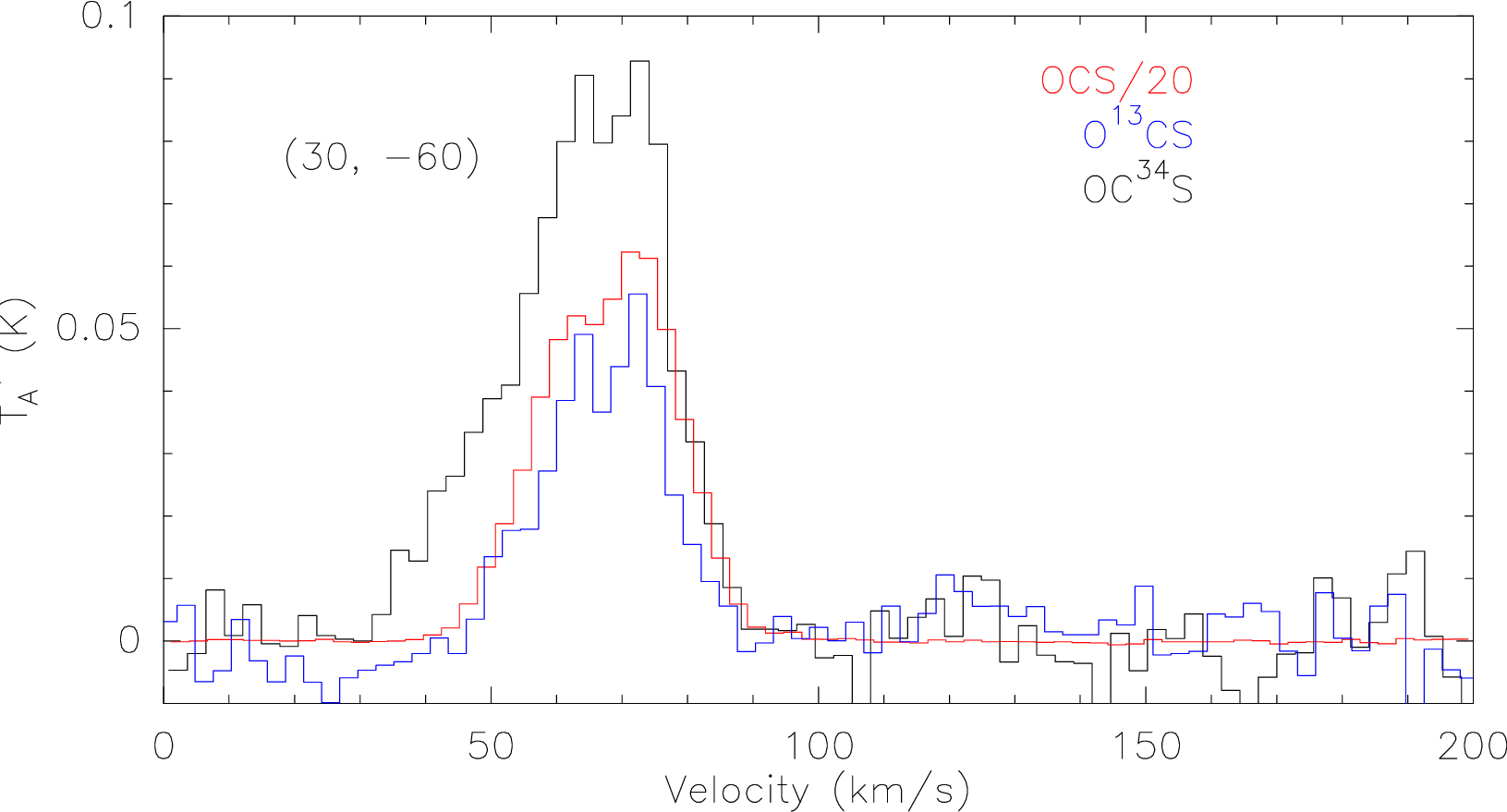}
\includegraphics[width=0.45\textwidth]{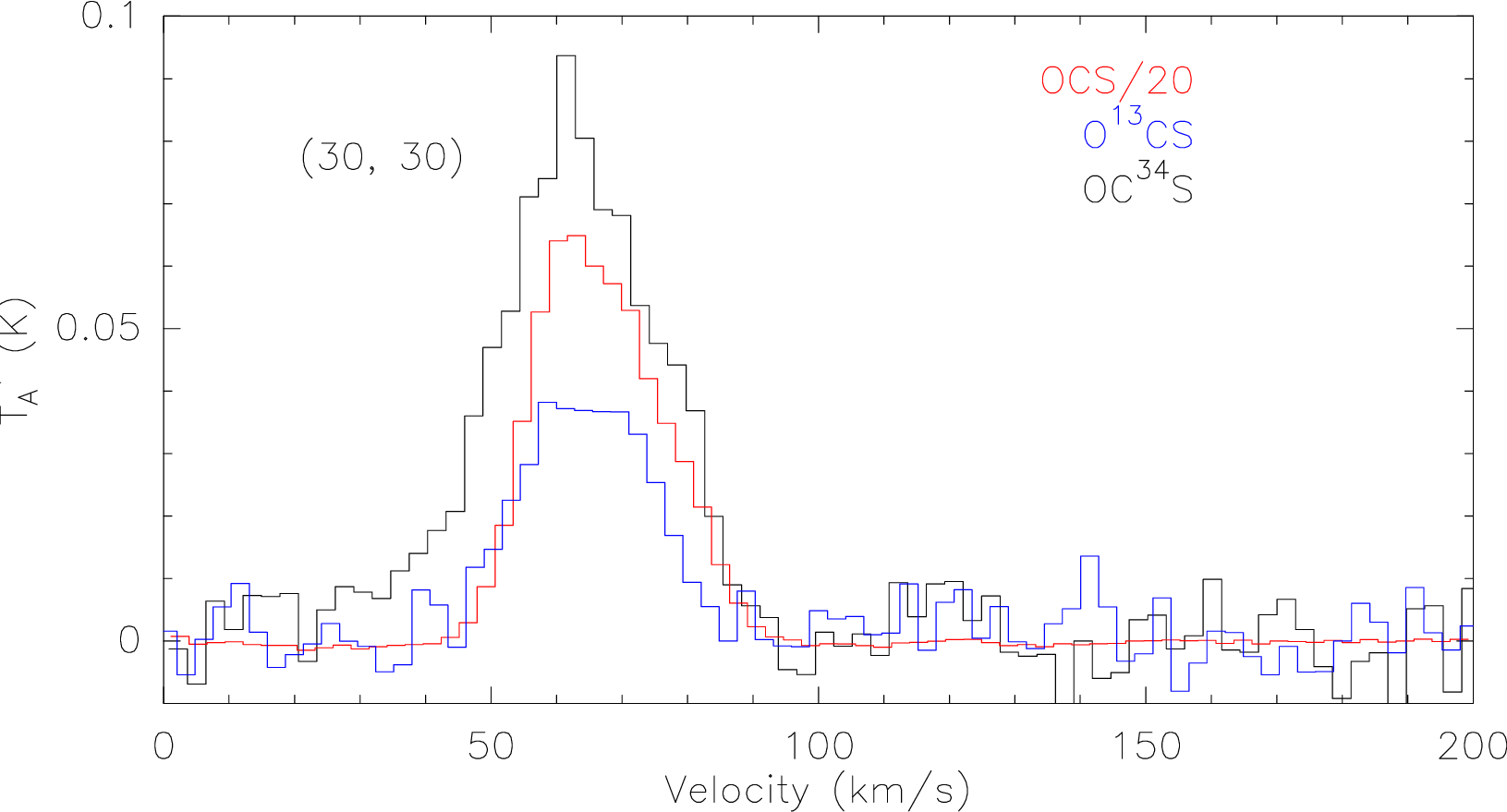}
\includegraphics[width=0.45\textwidth]{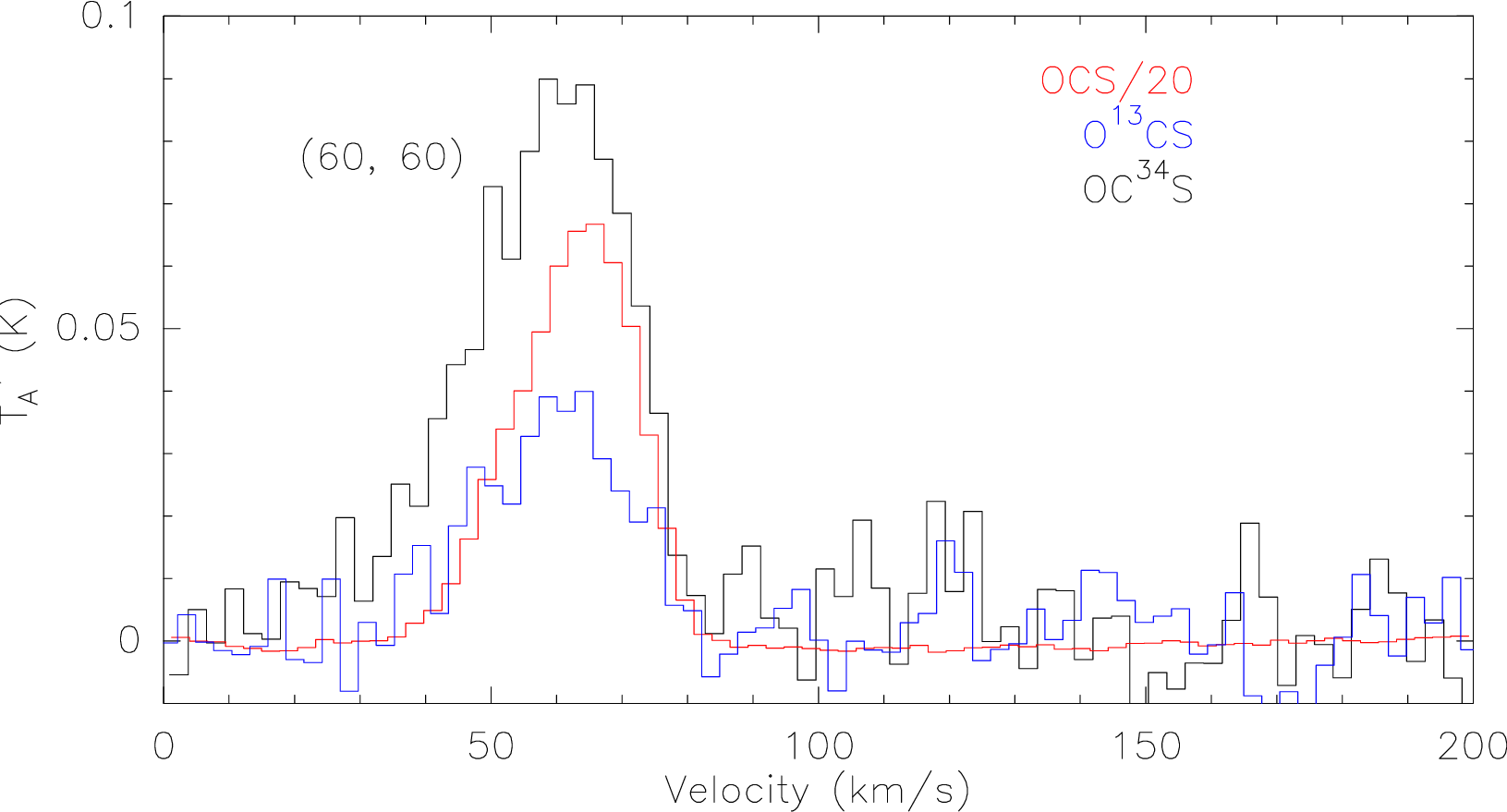}
\includegraphics[width=0.45\textwidth]{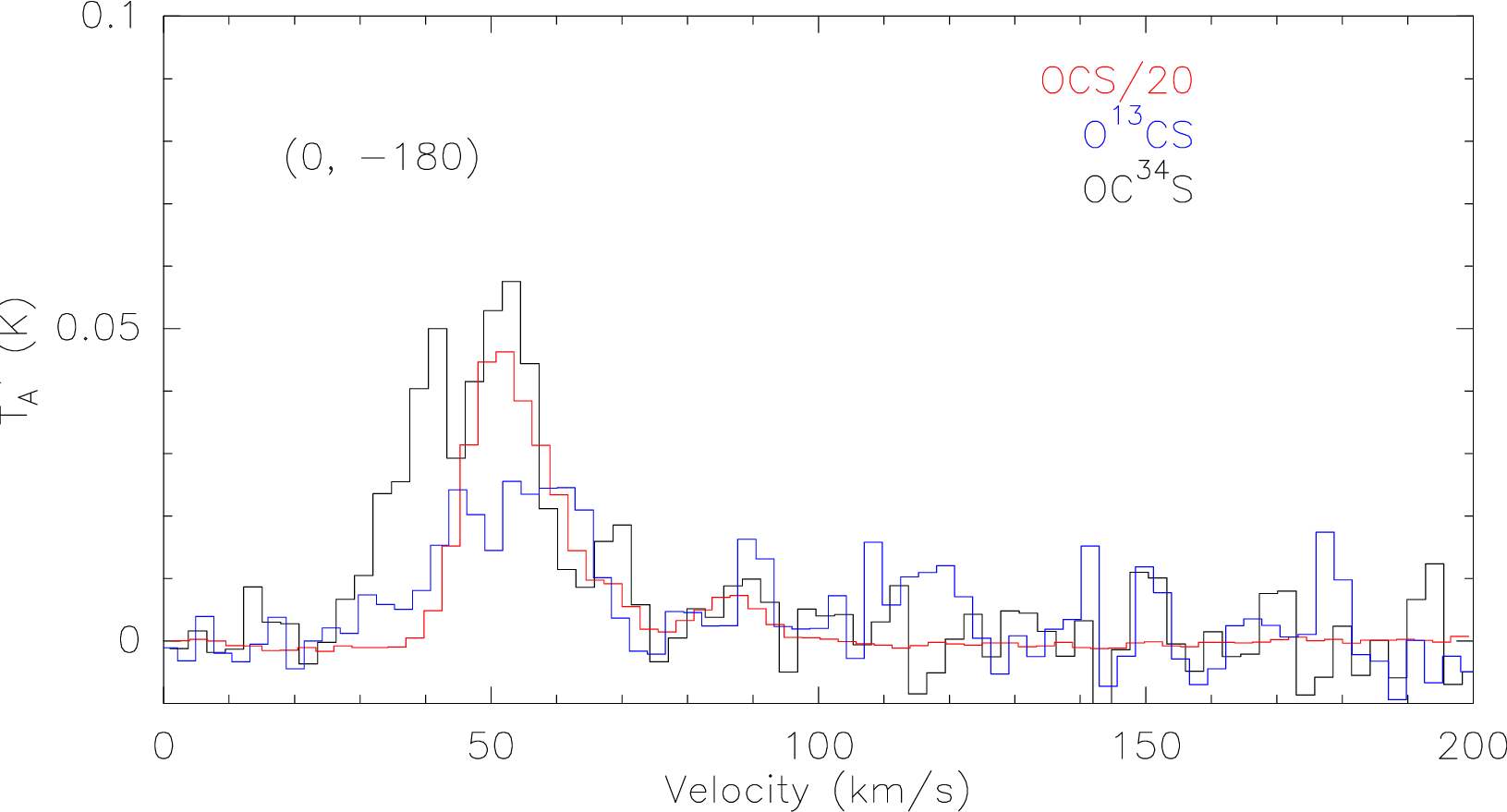}
\caption{The spectra of OCS 7-6 (red), O$^{13}$CS 7-6 (blue), and OC$^{34}$S 7-6 (black) toward positions in which OC$^{34}$S 7-6 line blend with NH$_2$D. The offsets relative to Sgr B2(N) are indicated at the left of each figure. }
\label{fig 1blend}
\end{figure*}

\begin{figure*}
\centering
\includegraphics[width=0.45\textwidth]{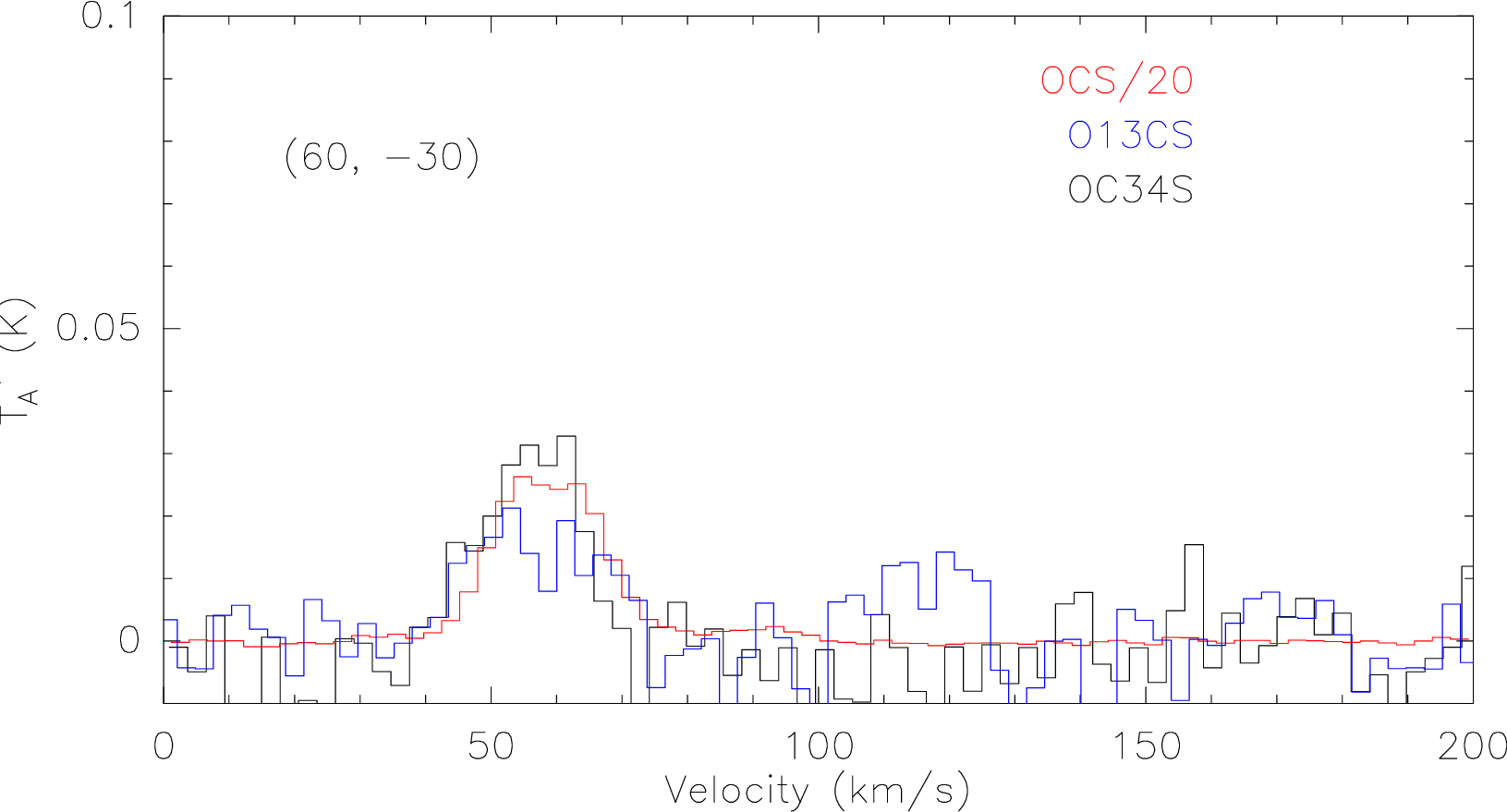}
\includegraphics[width=0.45\textwidth]{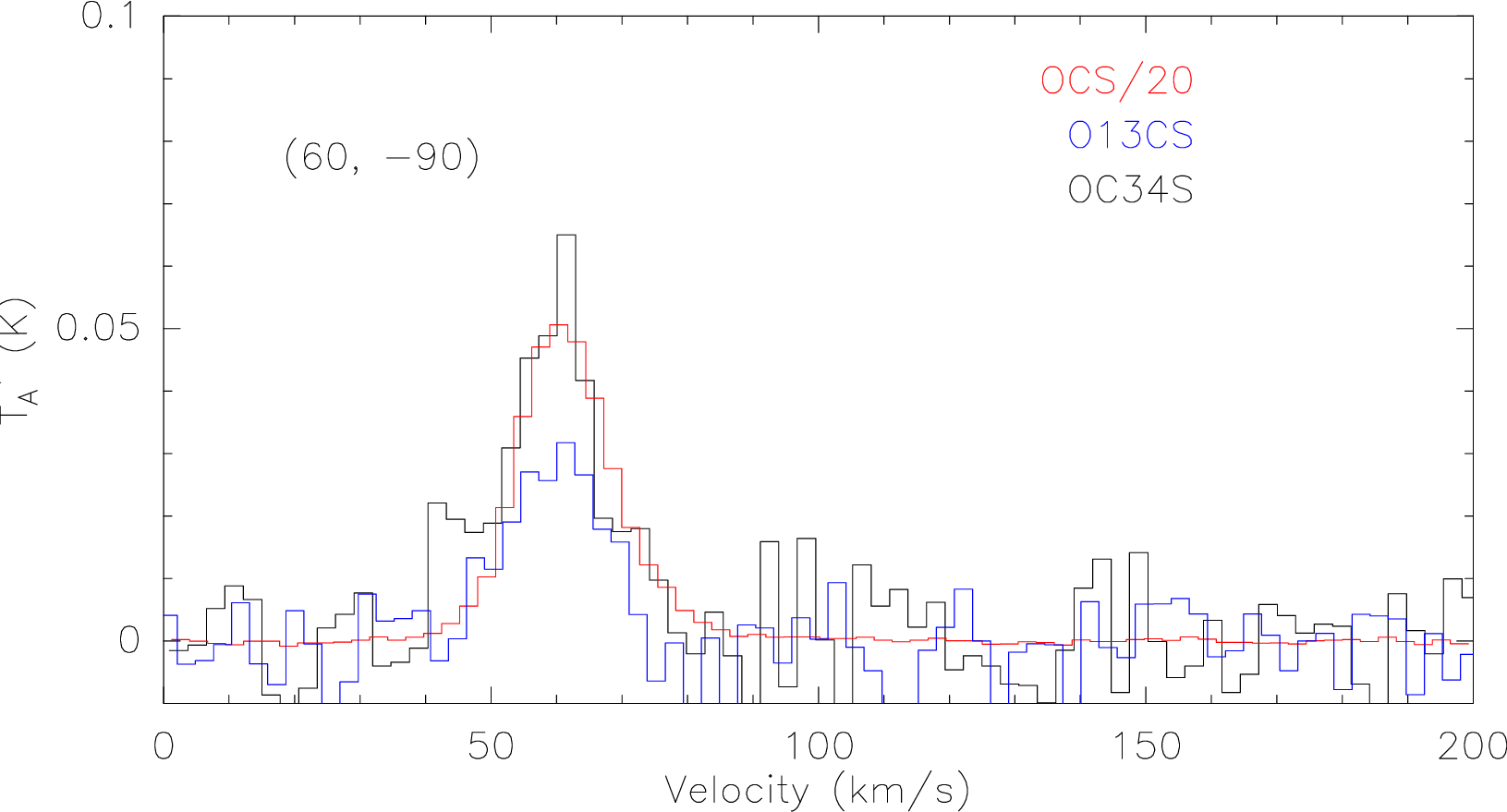}
\includegraphics[width=0.45\textwidth]{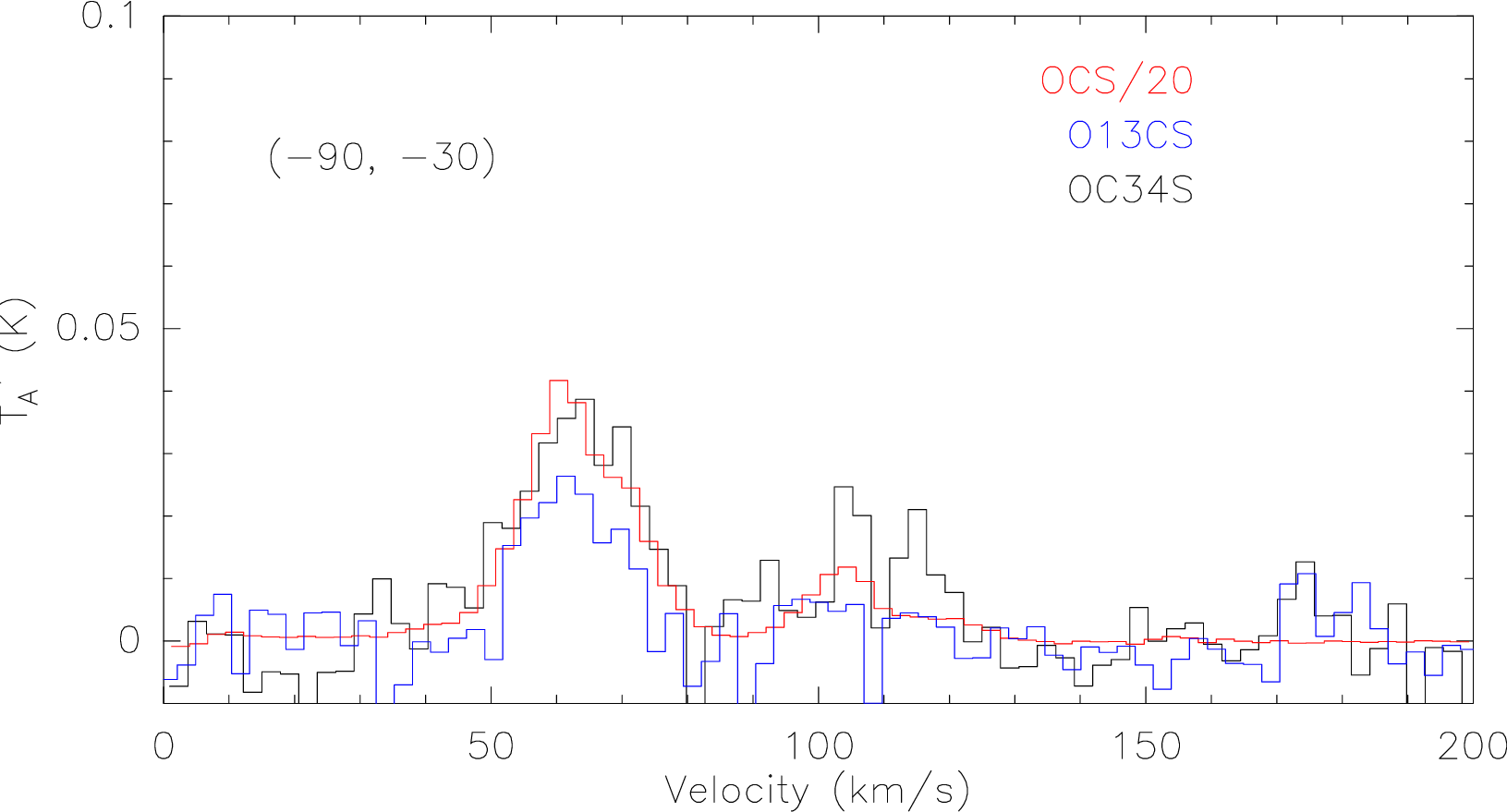}
\includegraphics[width=0.45\textwidth]{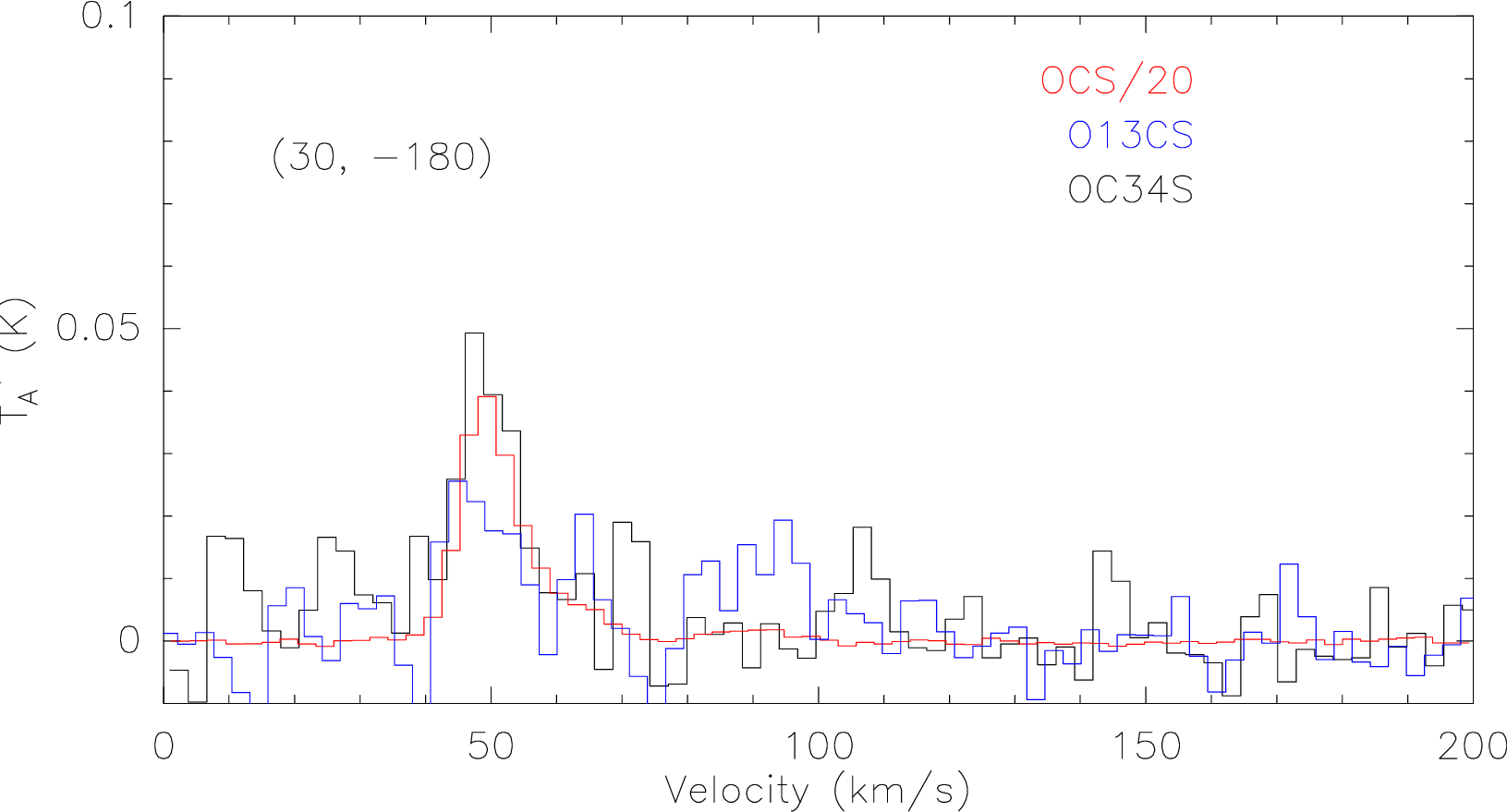}
\caption{The spectra of OCS 7-6 (red), O$^{13}$CS 7-6 (blue), and OC$^{34}$S 7-6 (black) toward selected positions in Sgr B2. The offsets relative to Sgr B2(N) are indicated at the left of each figure. }
\label{fig 1}
\end{figure*}

Figure \ref{map} presents the integrated intensity map of OCS 7-6 around Sgr B2. The mapping result is similar to that in \citet{jones08}. We could see from Figure \ref{map} that the OCS 7-6 emission peaks in Sgr B2(N). OCS is usually optically thick in massive star-forming regions \citep{li15}. The determination of sulfur isotope ratios with OCS and its isotopes may be affected by optical depth of OCS lines. By comparing with O$^{13}$CS lines, it is found that OCS lines are optically thick toward some positions. As mentioned above, NH$_2$D emission at 83060.244 MHz could affect the determination of integrated intensity of OC$^{34}$S. Figure \ref{fig 1blend} shows positions in which OC$^{34}$S 7-6 blend with NH$_2$D in the blue-shifted side. The possible contaminant lines are in the blue-shifted side of OC$^{34}$S 7-6, as is shown by the inconsistency between the OC$^{34}$S and other two transitions in the blue-shifted side. For example, the spectra toward (60, 60) and (0, -180) pointings have secondary peaks at lower velocities from their respective primary peaks. As the NH$_2$D transition has a higher frequency than OC$^{34}$S 7-6, it should affect the blue-shifted side of the spectra. Such contamination is negligible in Figure \ref{fig 1}. The judgement of the presence/absence of the contamination is done by eye-inspection of the emission excess or the secondary peak in the blue-shifted side. To solve the above problems related to the optically thickness and the spectral contamination, we first inspected the spectra of OC$^{34}$S 7-6 and OCS 7-6 carefully, and selected positions where the line profiles OC$^{34}$S 7-6 were consistent with those of OCS 7-6. These positions were regarded to be unaffected by NH$_2$D. The optical depth of OCS toward these positions were then calculated by comparing with O$^{13}$CS. Only positions with optically thin OCS lines will be used to determine the $^{32}$S/$^{34}$S ratio. 

 We estimated the optical depth of OCS J=7-6 line from:
\begin{equation}
\frac{T_{A}^*(O^{12}CS)}{T_{A}^*(O^{13}CS)} \sim \frac{1-e^{-\tau (O^{12}CS)}}{1-e^{-\tau (O^{12}CS)/R_C}}, R_{C}= \frac{^{12}C}{^{13}C},
\end{equation}
in which $T_A^*$ is the peak main brightness temperature. We adopt the $^{12}C/^{13}C$ ratio of 31, which is derived from NH$_2$CHO and NH$_2^{13}$CHO (Zheng et al., submitted.). It is found that
$\frac{T_{A}^*(O^{12}CS)}{T_{A}^*(O^{13}CS)}$ is larger than 31 toward some pointings, in which $\tau$ was unmeasurable. OCS emission was regarded to be optically thin for these positions.

At last, four positions with optically thin OCS lines, where OC$^{34}$S 7-6 lines are not affected by NH$_2$D emission at 83060.244 MHz either, were selected to determine the sulfur isotope ratios in Sg B2. The green triangles denote these positions in Figure \ref{map}. All of these positions are far away from the central region of Sgr B2 complex. Table \ref{32s34s} lists the selected positions, and the physical parameters of lines toward these positions. Figure \ref{fig 1} presents the lines of OCS, O$^{13}$CS, and OC$^{34}$S 7-6 emission toward these positions.

\subsection{$^{32}$S/$^{34}$S ratio derived directly from OCS and OC$^{34}$S }

\begin{figure*}
\centering
\includegraphics[width=0.45\textwidth]{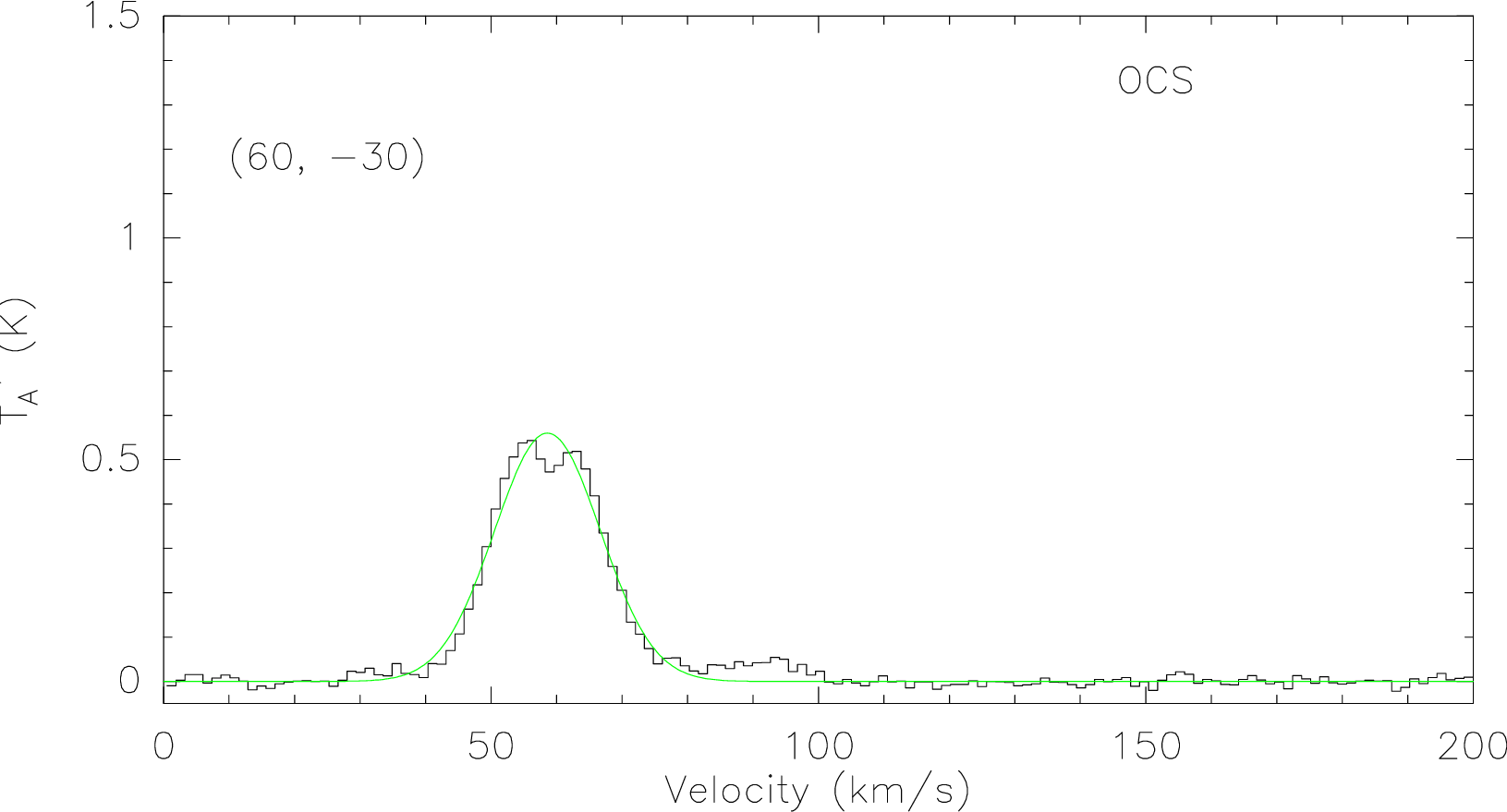}
\includegraphics[width=0.45\textwidth]{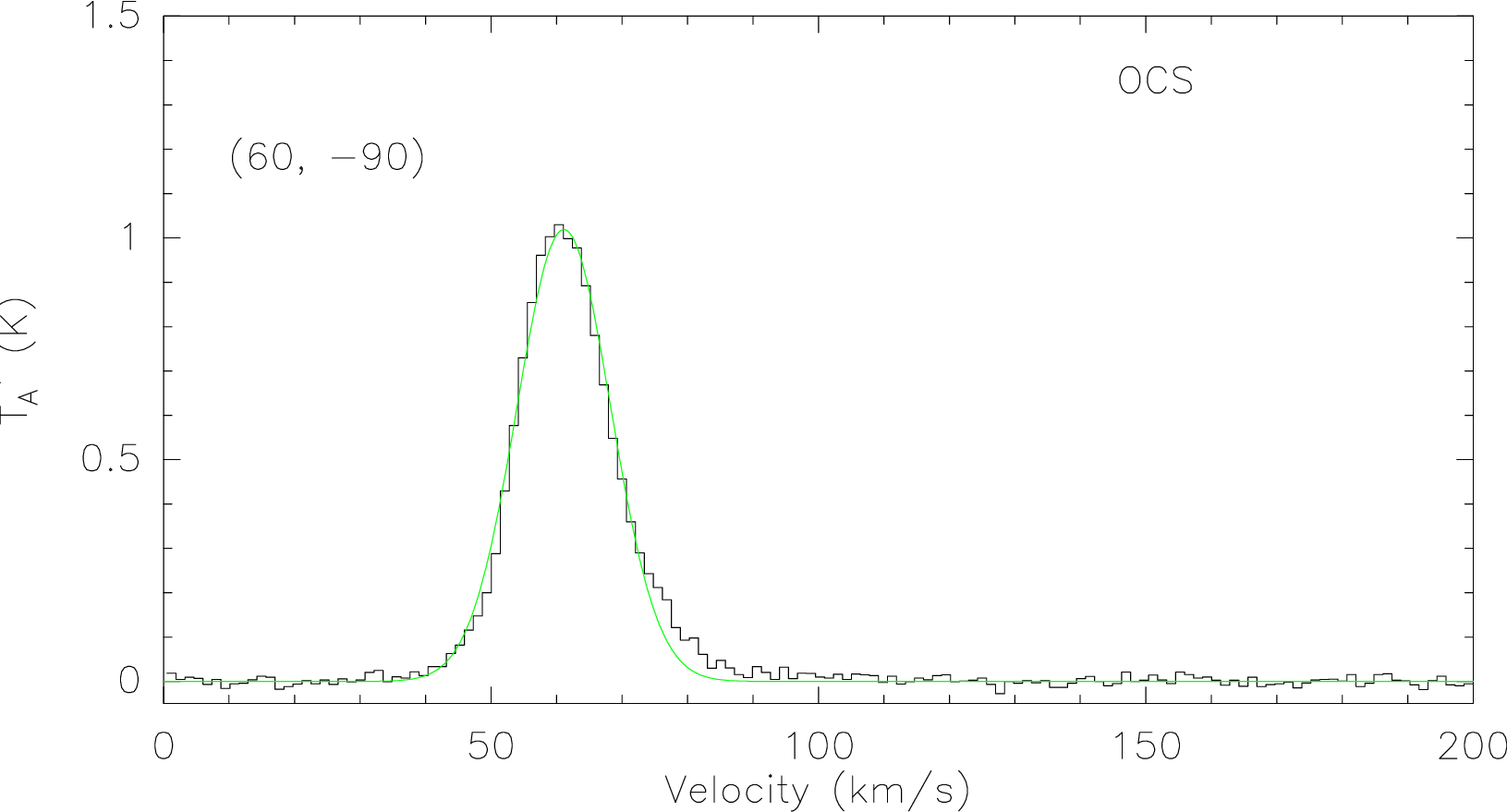}
\includegraphics[width=0.45\textwidth]{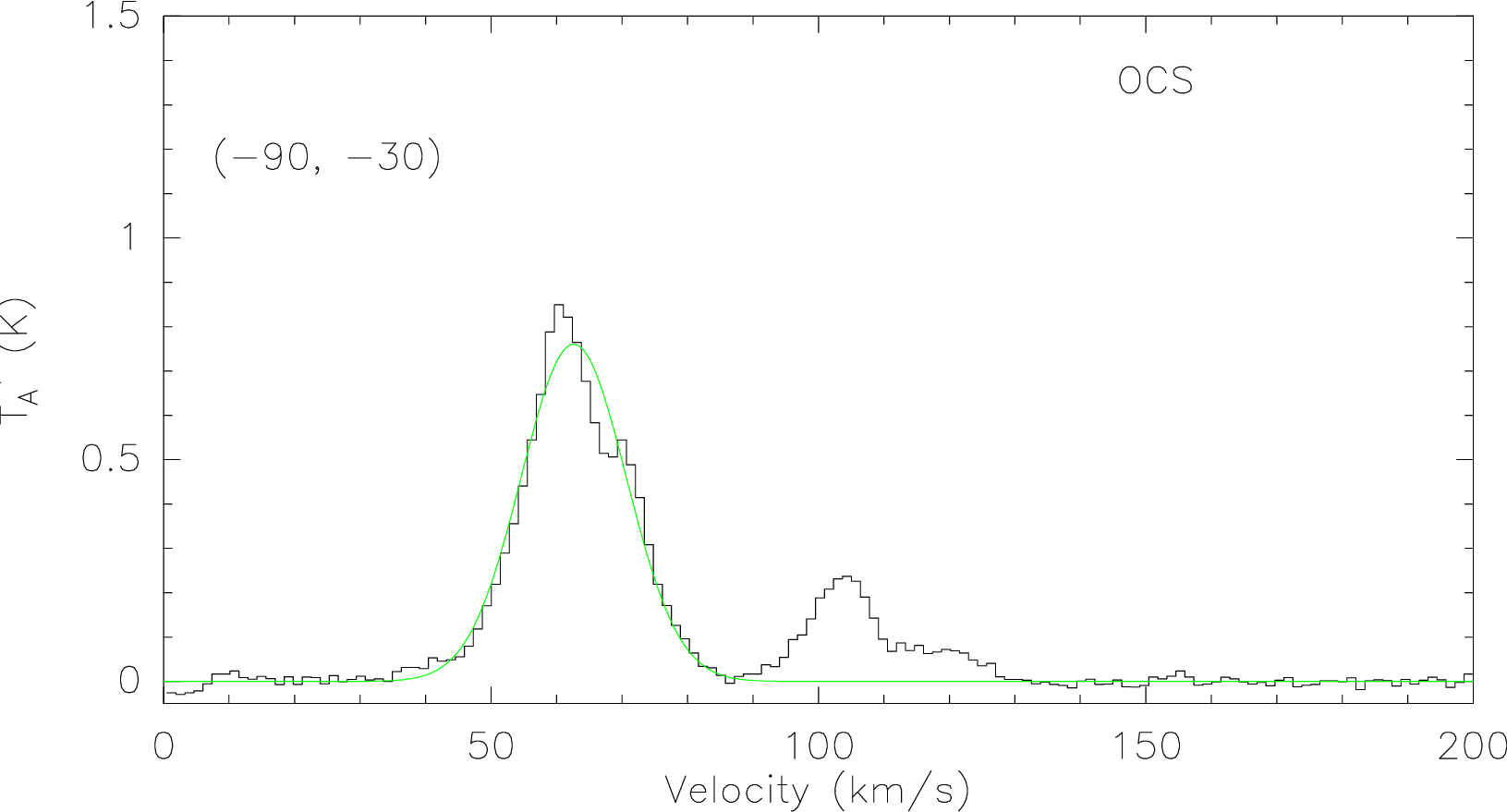}
\includegraphics[width=0.45\textwidth]{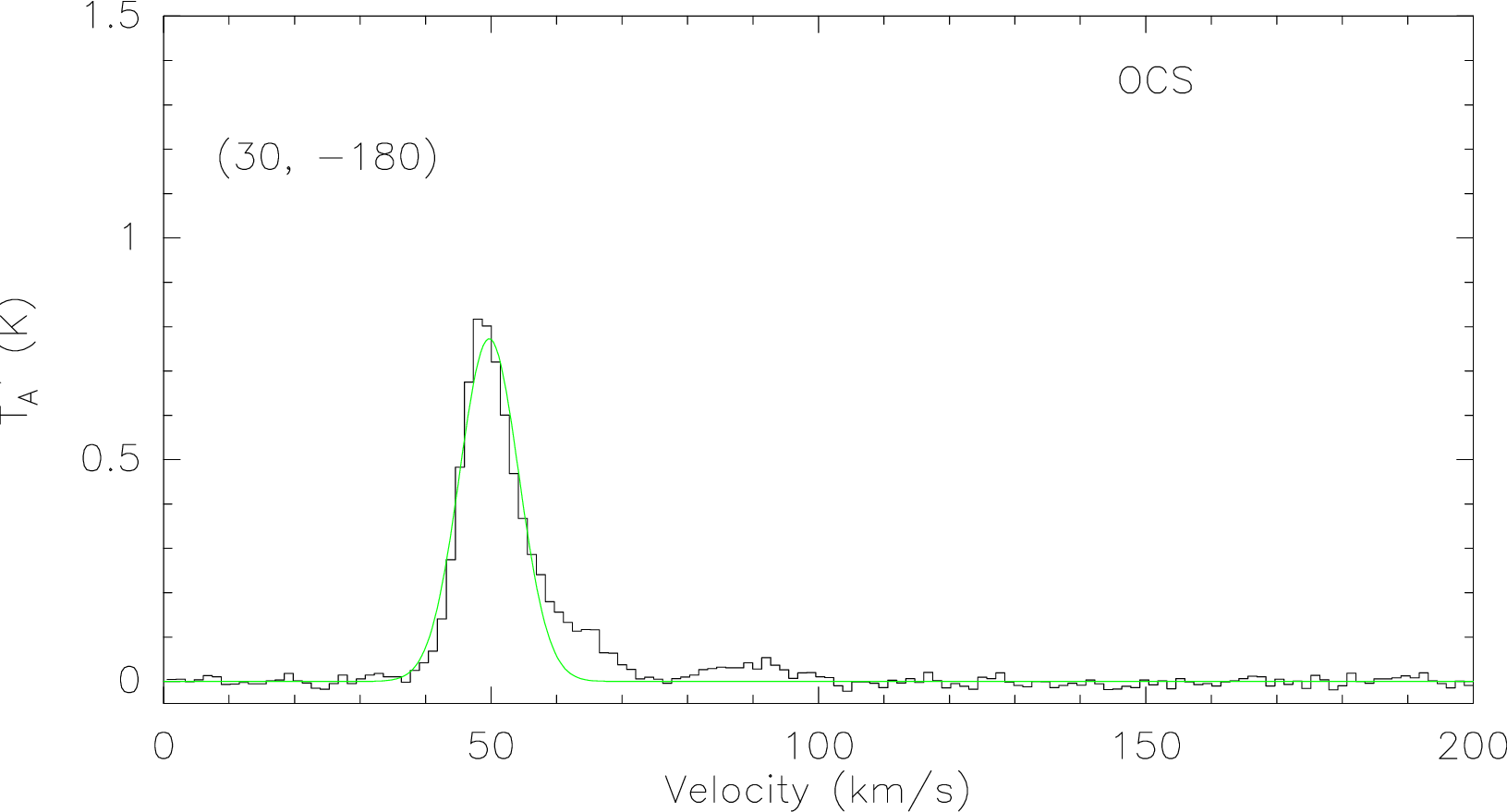}
\caption{The spectra of OCS 7-6 toward selected positions in Sgr B2. The offsets relative to Sgr B2(N) are indicated at the left of each figure. The gaussian fitting results are shown in green. }
\label{ocsfit}
\end{figure*}

\begin{figure*}
\centering
\includegraphics[width=0.45\textwidth]{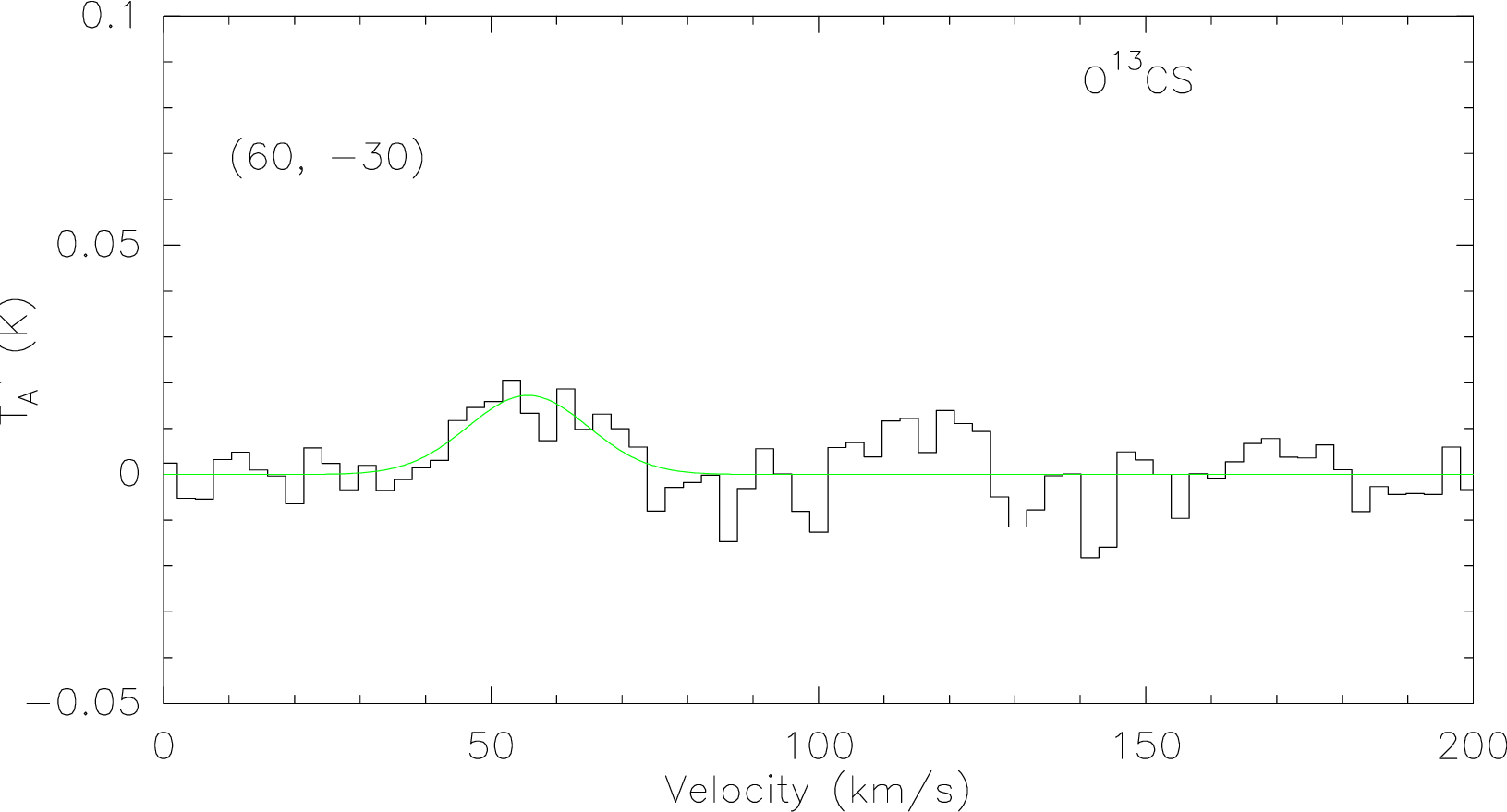}
\includegraphics[width=0.45\textwidth]{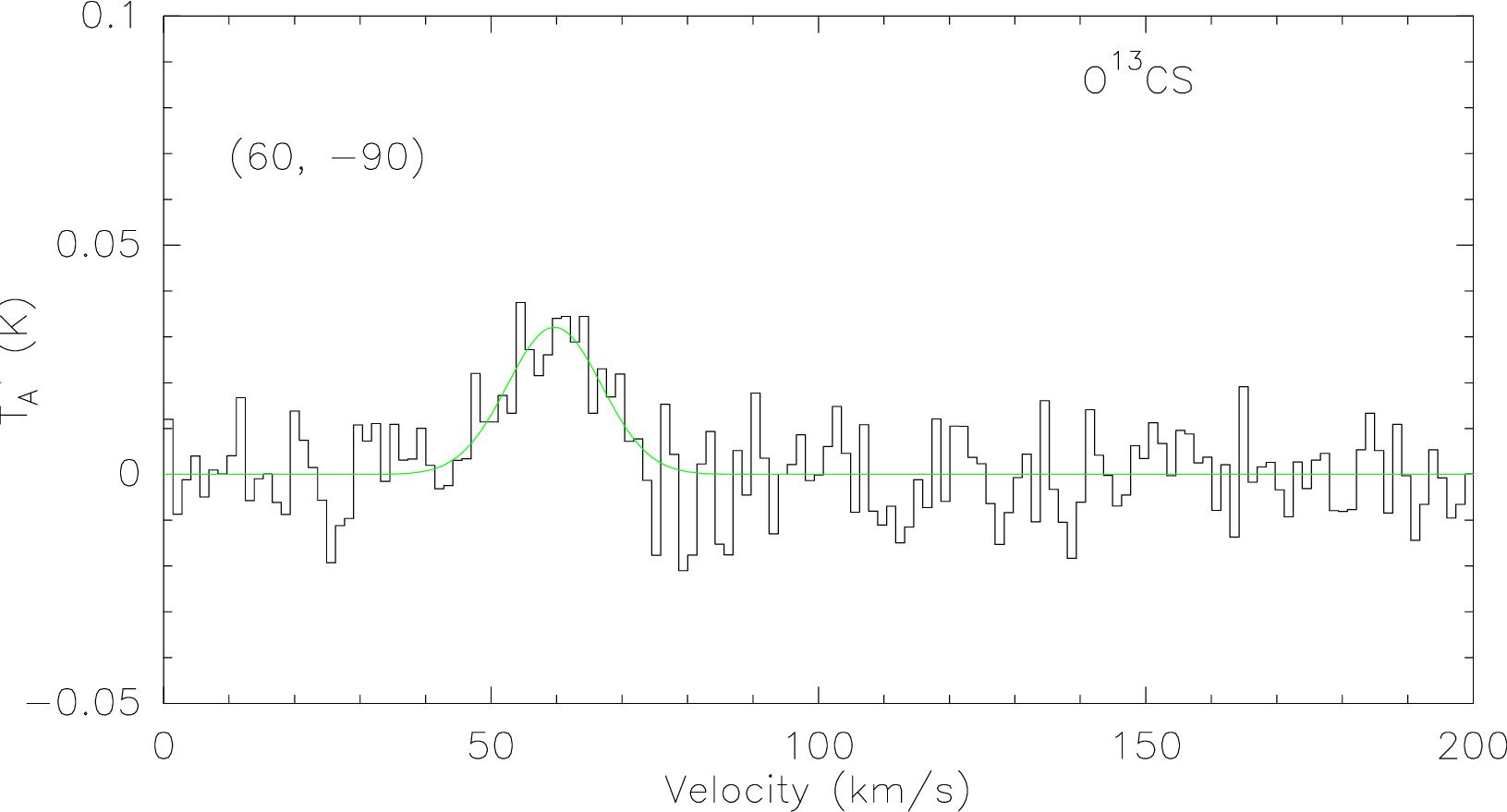}
\includegraphics[width=0.45\textwidth]{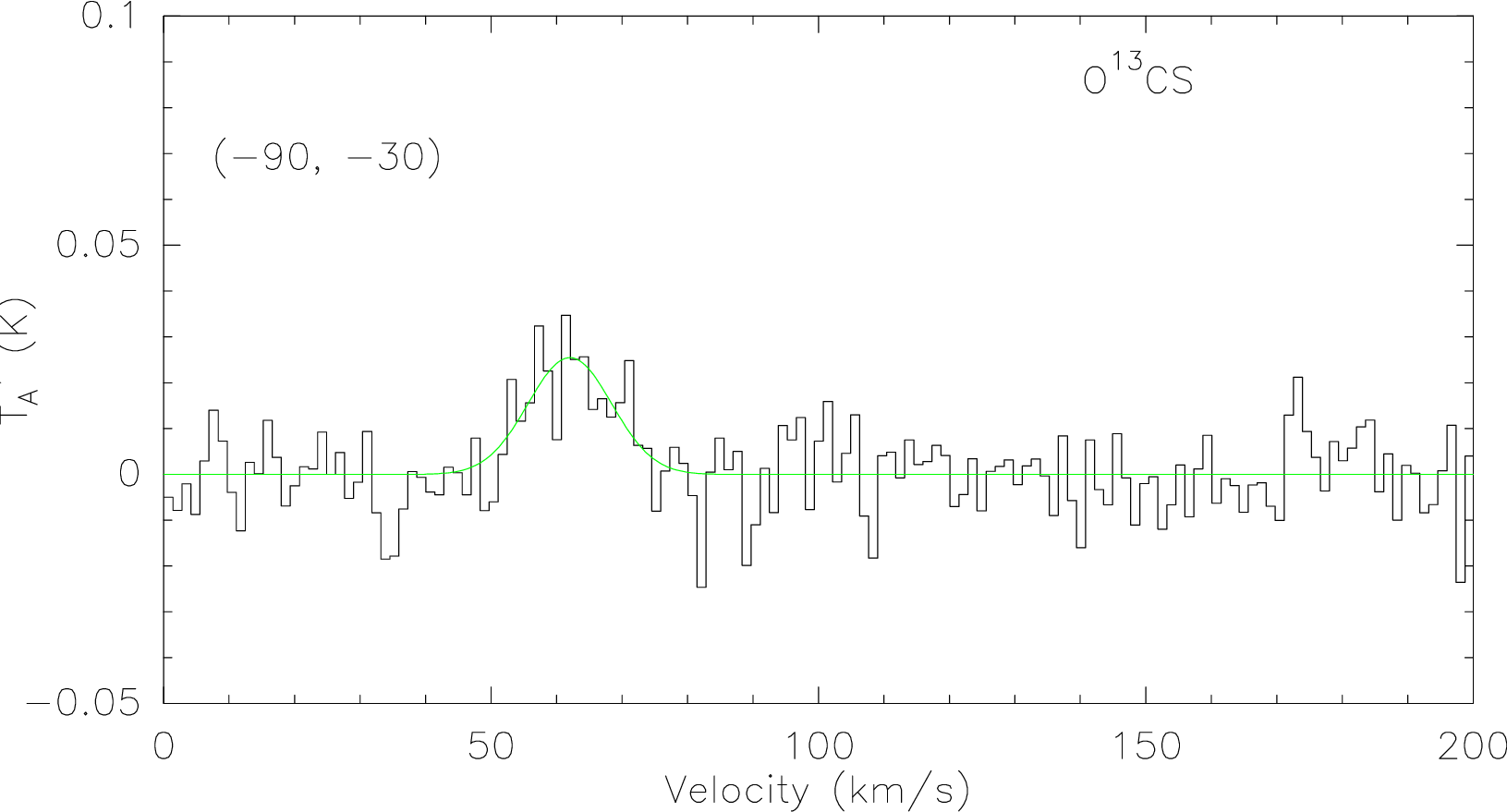}
\includegraphics[width=0.45\textwidth]{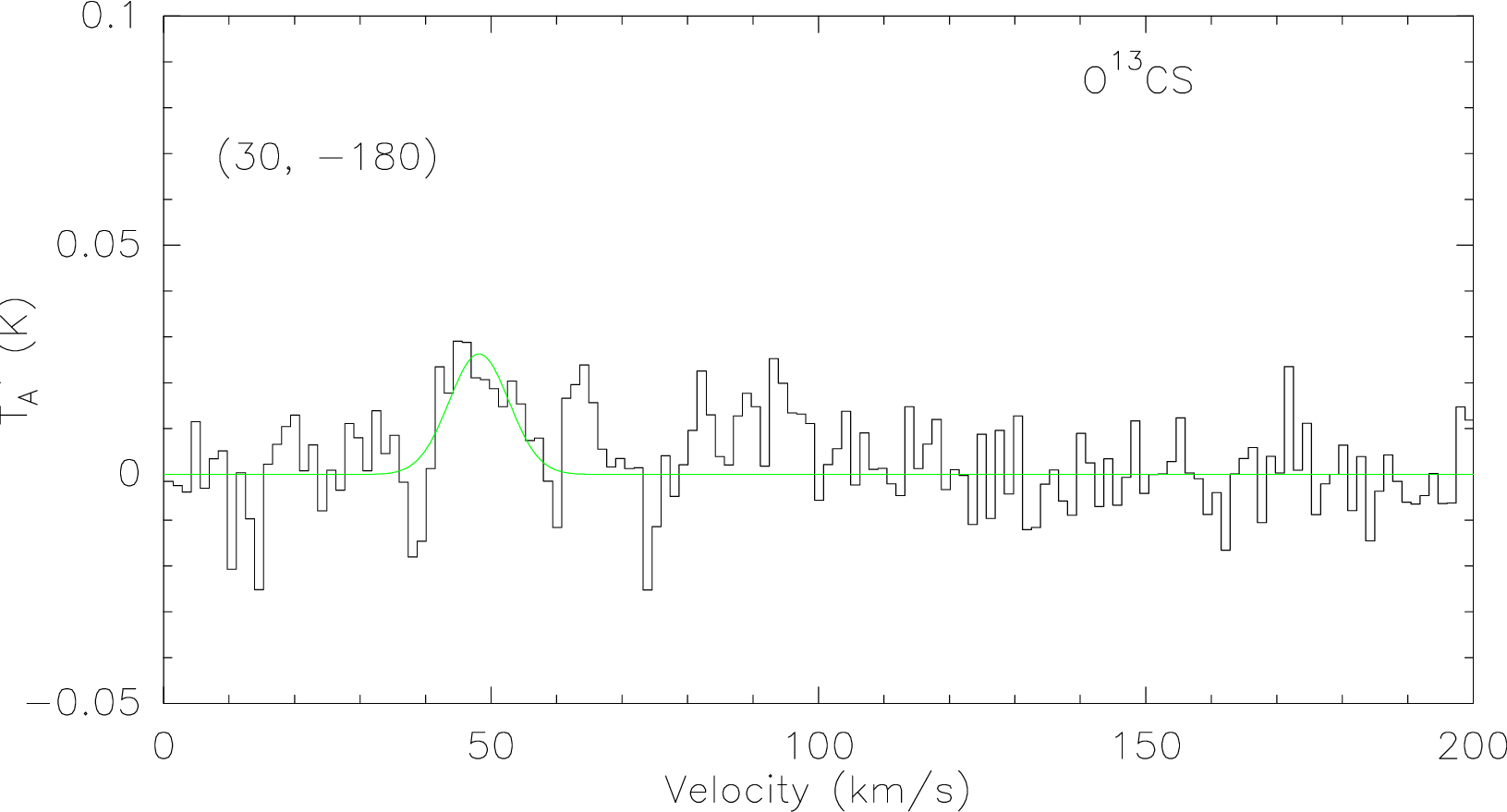}
\caption{The spectra of O$^{13}$CS 7-6 toward selected positions in Sgr B2. The offsets relative to Sgr B2(N) are indicated at the left of each figure. The gaussian fitting results are shown in green. For O$^{13}$CS toward (60, -30), the velocity resolution was smoothed to 2.8 km s$^{-1}$ to improve the signal-to-noise ratio.}
\label{o13csfit}
\end{figure*}

\begin{figure*}
\centering
\includegraphics[width=0.45\textwidth]{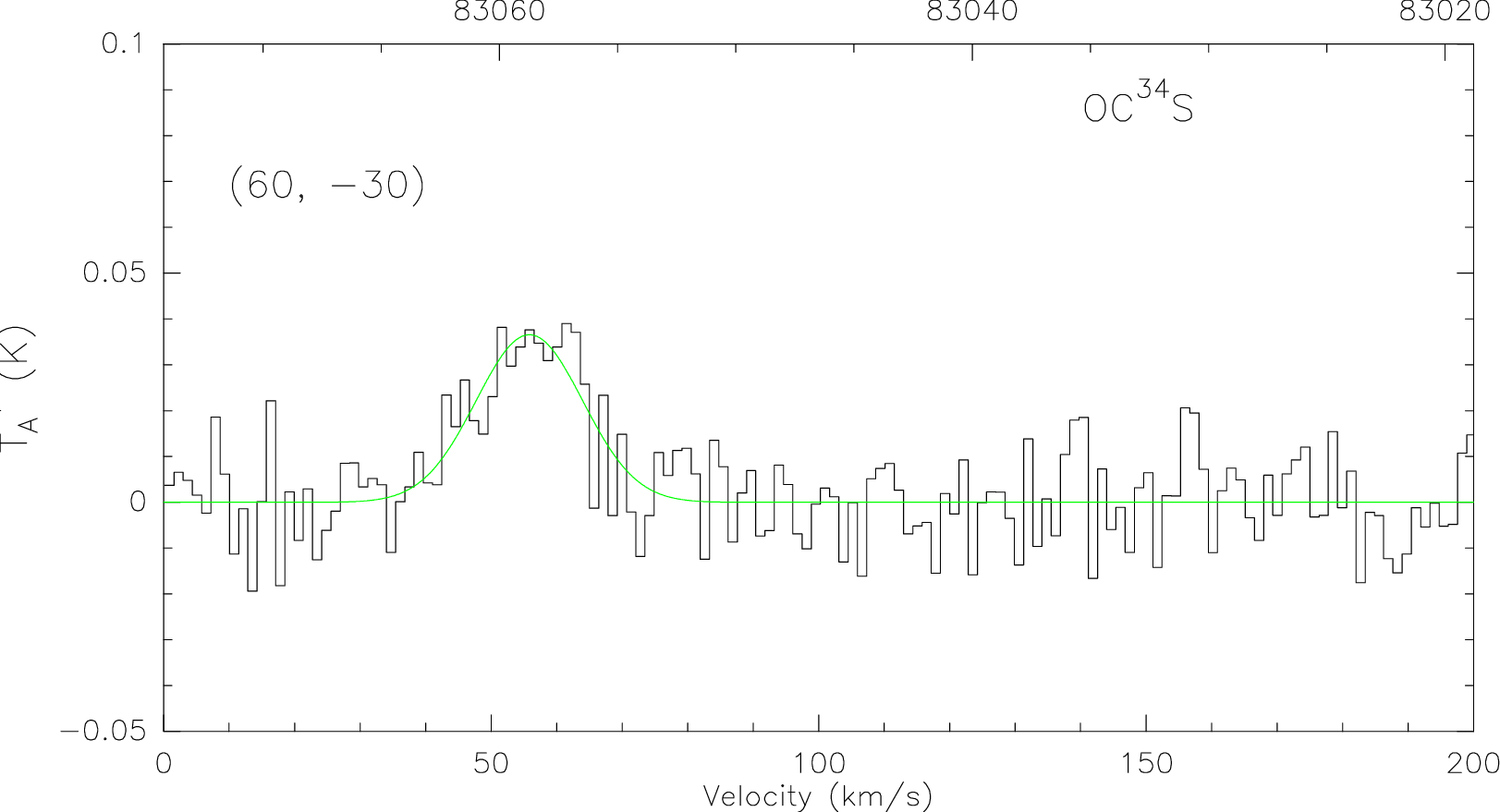}
\includegraphics[width=0.45\textwidth]{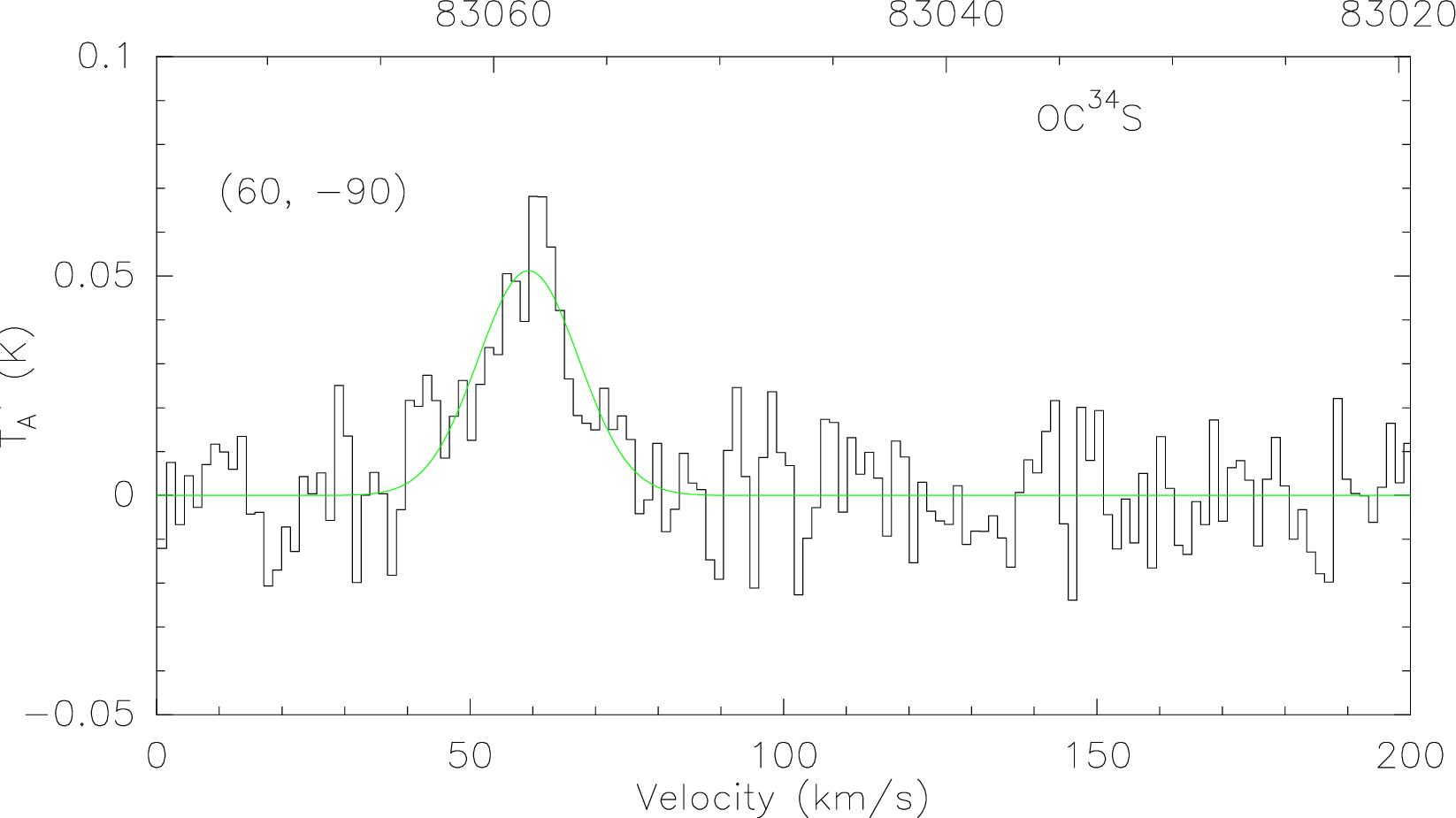}
\includegraphics[width=0.45\textwidth]{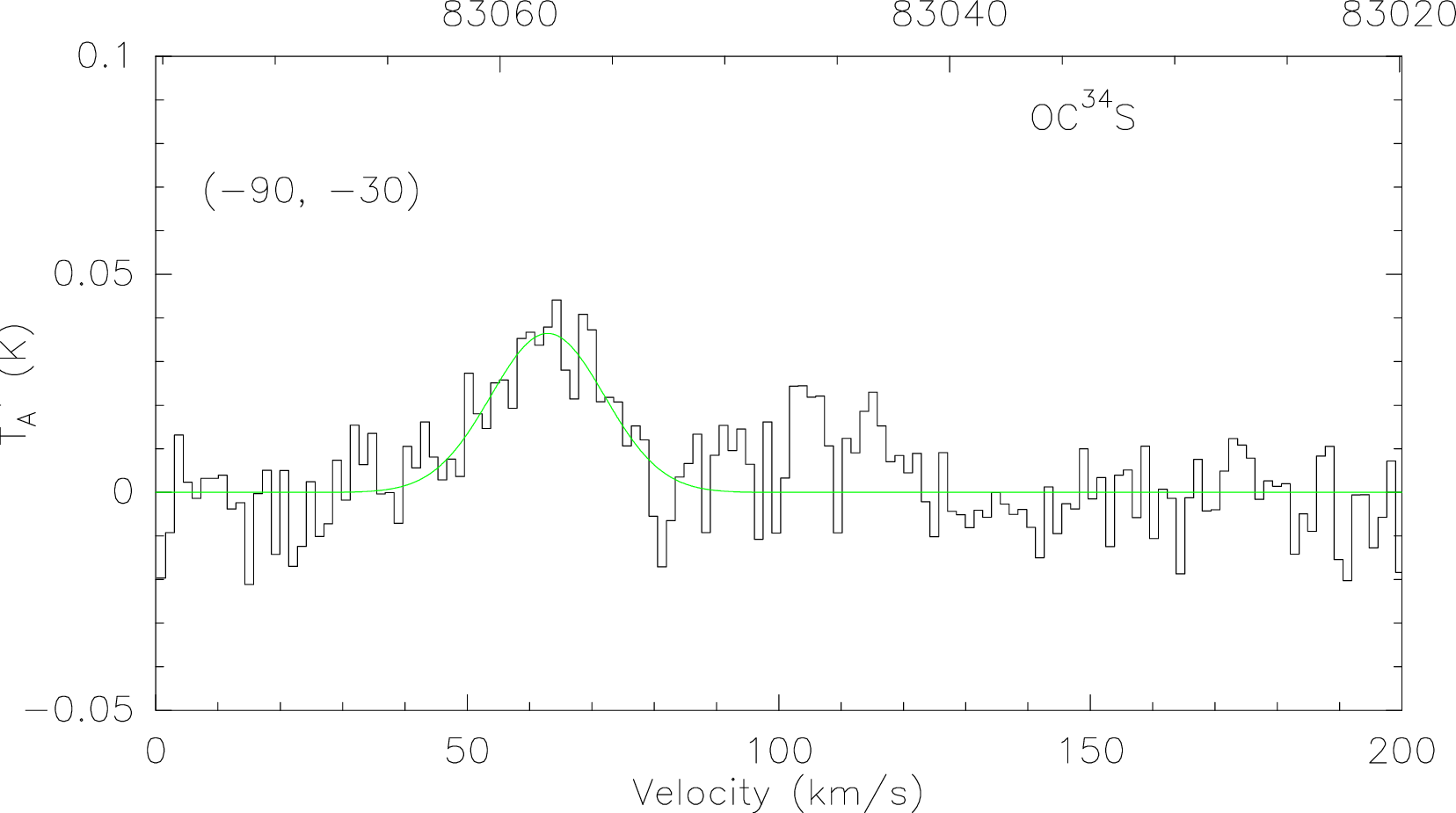}
\includegraphics[width=0.45\textwidth]{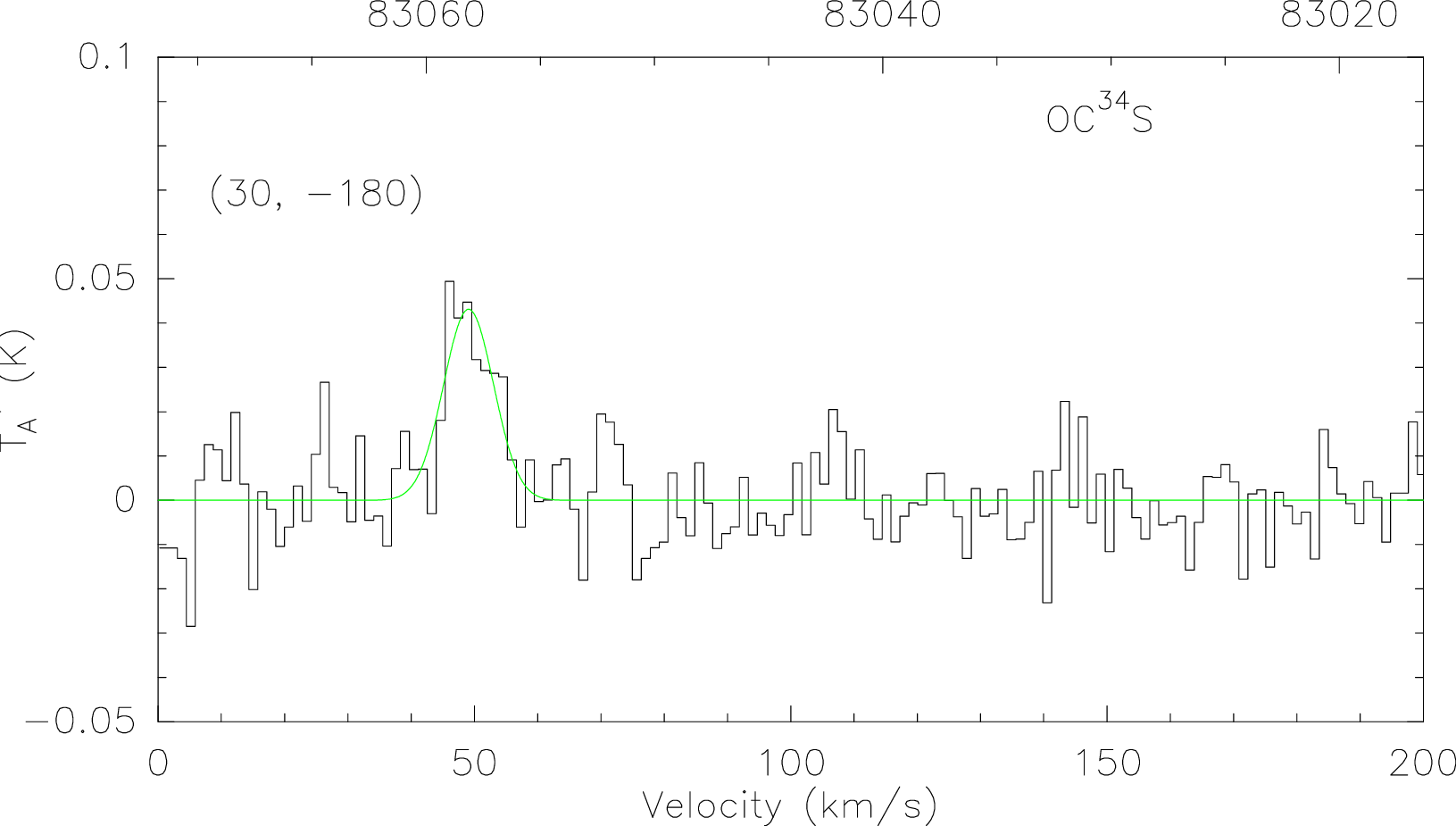}
\caption{The spectra of OC$^{34}$S 7-6 toward selected positions in Sgr B2. The offsets relative to Sgr B2(N) are indicated at the left of each figure. The gaussian fitting results are shown in green. }
\label{oc34sfit}
\end{figure*}

Figure \ref{ocsfit}, \ref{o13csfit}, \ref{oc34sfit} present the gaussian fitting results of OCS, O$^{13}$CS and OC$^{34}$S toward selected positions. For a given transition the frequencies, upper level energies and Einstein coefficients for the three isotopologues are comparable; therefore, the abundance ratio can be directly derived from the line ratios. Table \ref{32s34s} presents the Gaussian fitting results of OCS and OC$^{34}$S toward four positions and the velocity-corrected averaged spectra of these four positions. The line parameters, including the antenna temperature, centroid velocity, FWHM, and integrated intensities are obtained by gaussian fitting. The $^{32}$S/$^{34}$S ratios derived from the integrated intensity ratios of OCS and OC$^{34}$S: 
\begin{equation}
\frac{^{32}S}{^{34}S}  \sim \frac {I(OCS) } {I(OC^{34}S) },
\end{equation}
which requires that OCS and OC$^{34}$S are optically thin. The calculated values range from 15.1 to 20.3, which are consistent within the uncertainties with each other. 

\begin{figure*}
\centering
\includegraphics[width=0.45\textwidth]{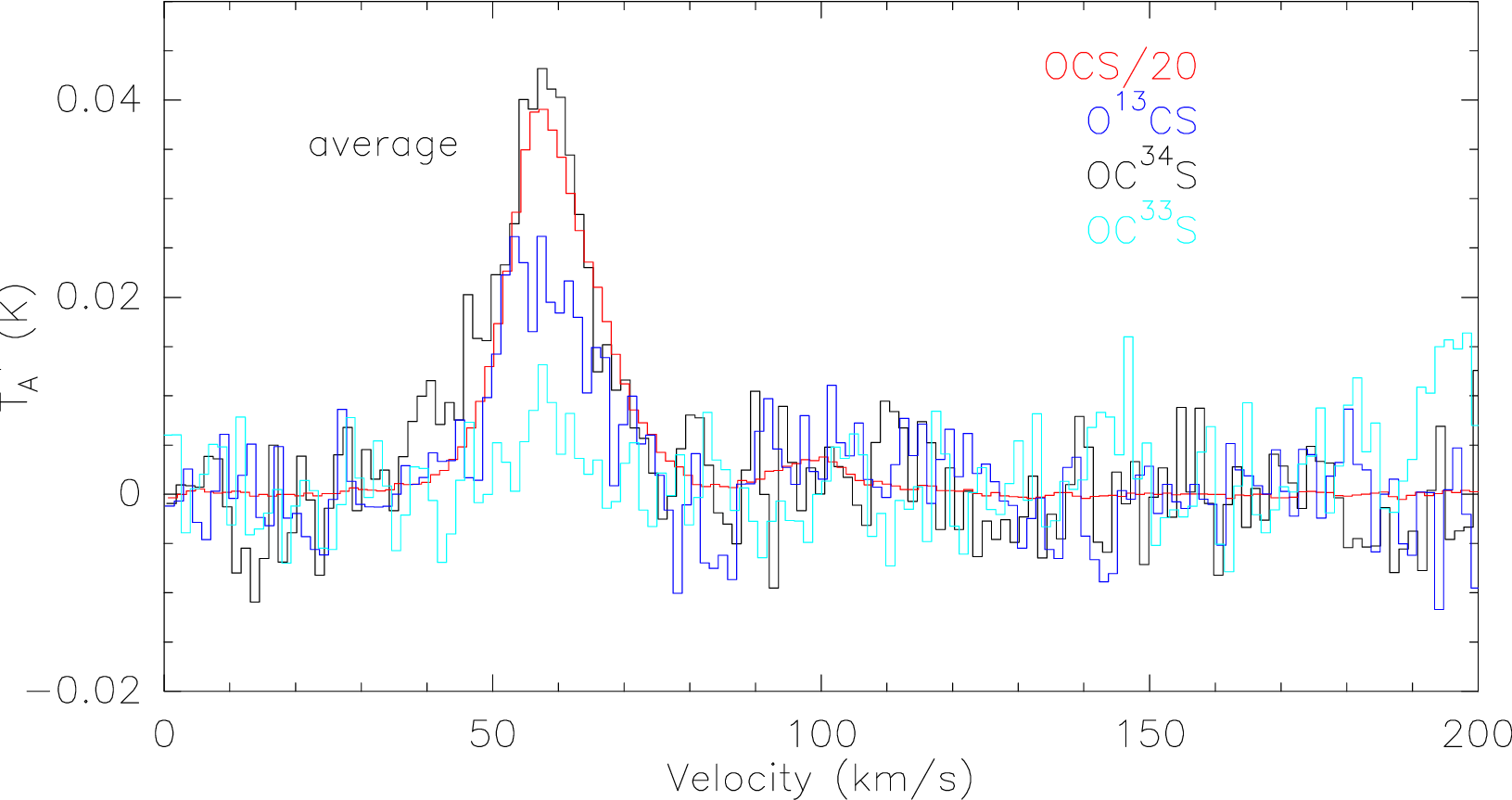}
\includegraphics[width=0.45\textwidth]{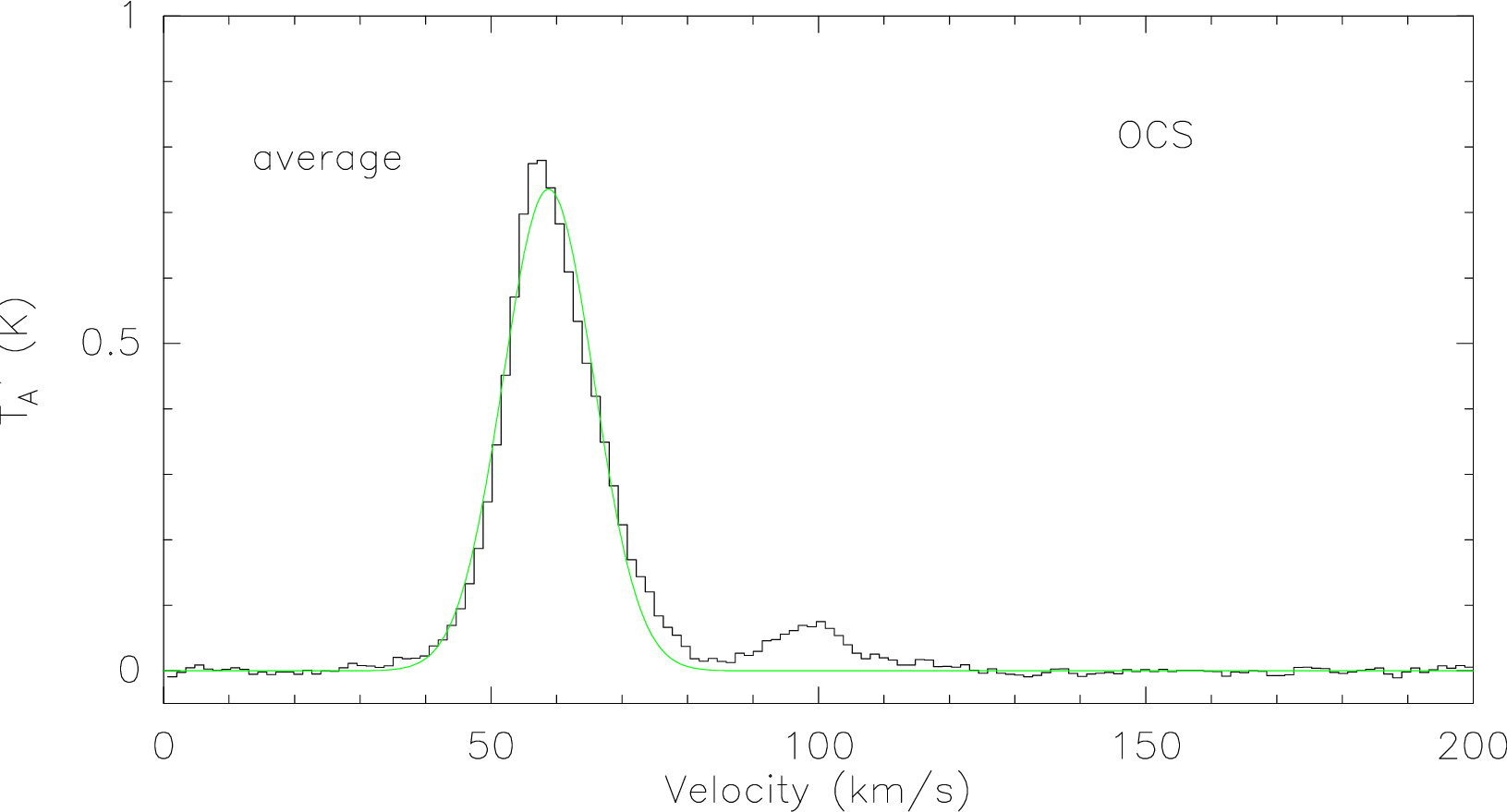}
\includegraphics[width=0.45\textwidth]{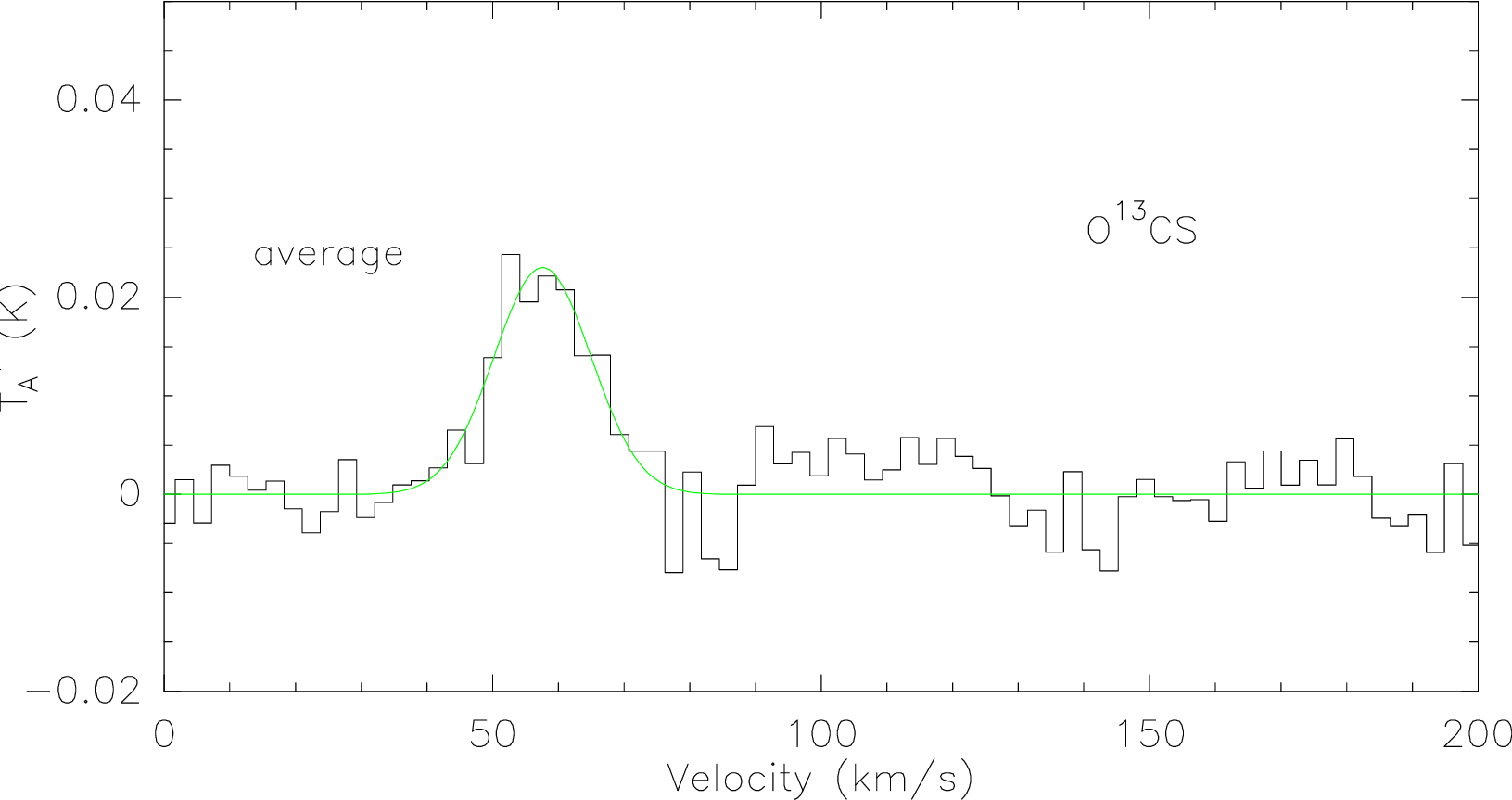}
\includegraphics[width=0.45\textwidth]{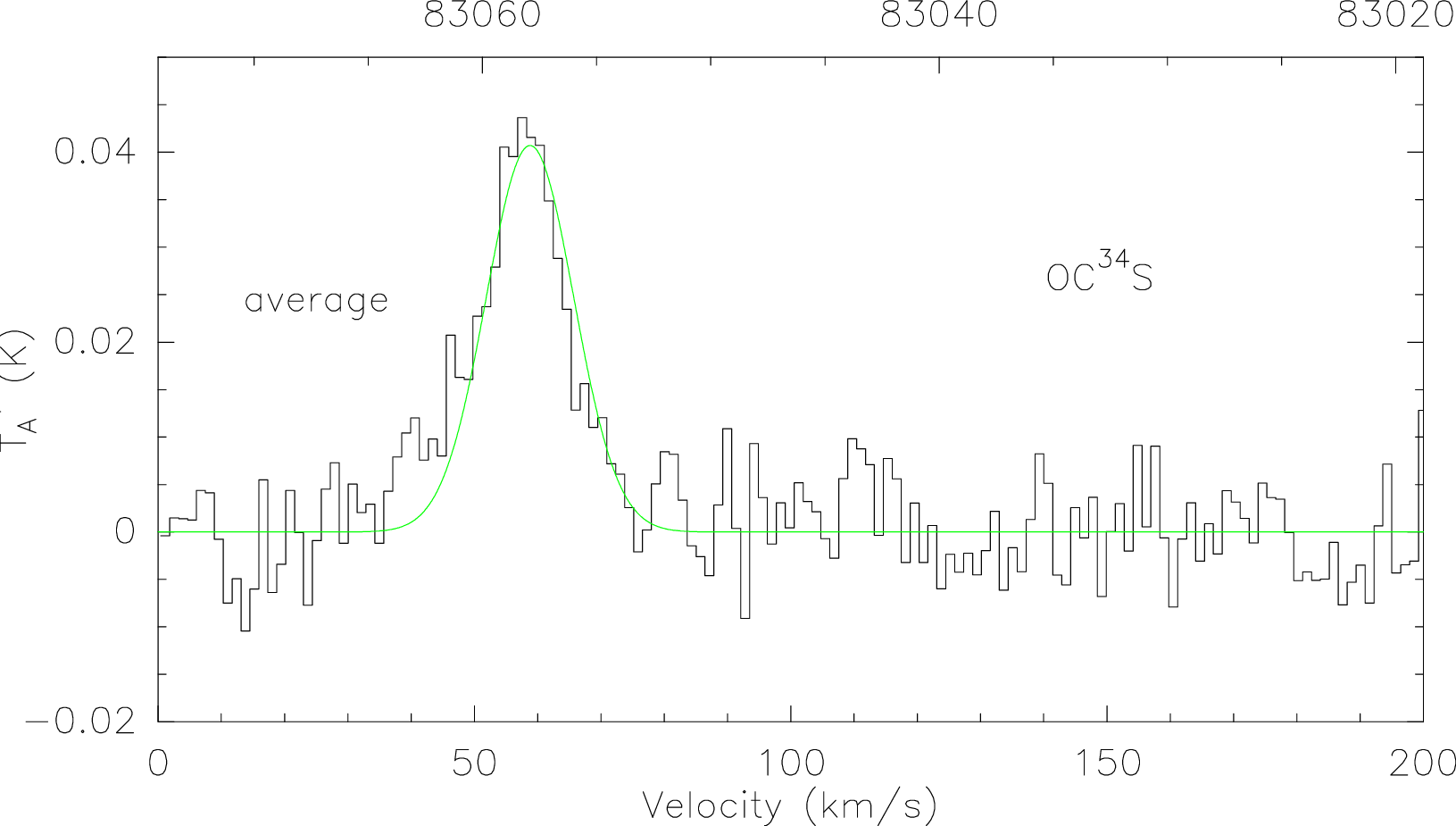}
\includegraphics[width=0.45\textwidth]{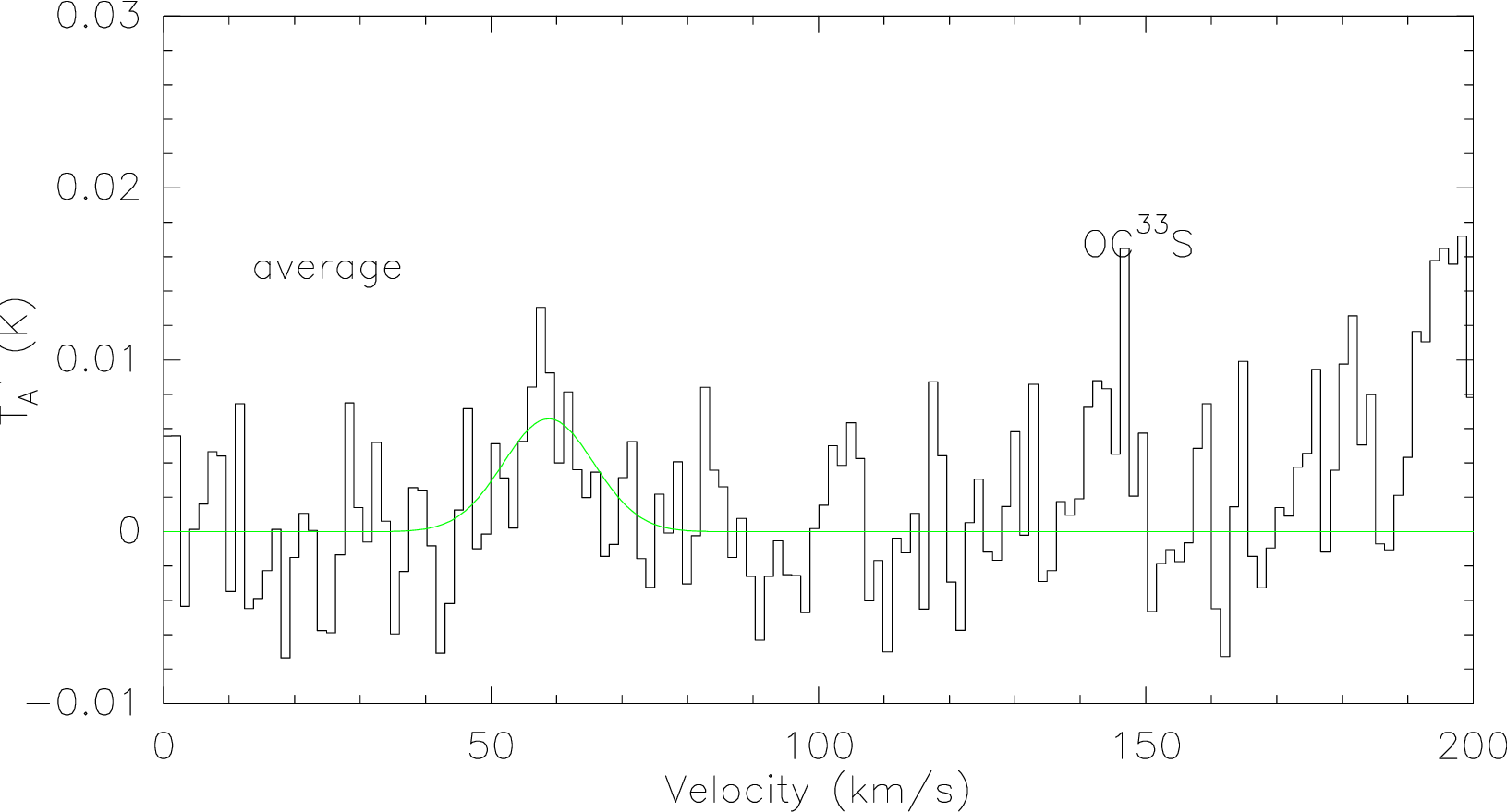}
\caption{The averaged spectra of OC$^{33}$S, OCS, O$^{13}$CS and OC$^{34}$S 7-6 toward selected positions in Sgr B2. The gaussian fitting results are shown in green. For O$^{13}$CS, the velocity resolution was smoothed to 2.8 km s$^{-1}$ to improve the signal-to-noise ratio.}
\label{oc33s}
\end{figure*}

We also averaged spectral lines toward selected positions in Table \ref{32s34s}. Figure \ref{oc33s} shows the velocity-corrected averaged spectra of OCS, O$^{13}$CS, OC$^{33}$S, and OC$^{34}$S. 
The system velocities toward each positions are different to each other,  with velocity difference up to 13 km s$^{-1}$. We used the velocity information of each positions (See Table  \ref{32s34s}) based on detected strong lines to do the alignment. Then these spectra were averaged by setting ``set align frequency", which means that spectra were averaged along the frequency axis. 
The $^{34}$S/$^{32}$S isotopic ratio is calculated to be 17.1$\pm$0.9, which is well consistent with those derived from the CS and its isotoplogues toward the +50 km s$^{-1}$ Cloud and envelope of Sgr B2(N) presented by \citet{humire20}, but is larger than those obtained in \citet{yu20} and \citet{armijos15}.

\subsection{$^{34}$S/$^{33}$S ratio derived directly from OC$^{34}$S and OC$^{33}$S}

OC$^{33}$S 7-6 was relatively weak. It was detected above 3$\sigma$ levels toward only a few positions. To improve the signal to noise ratio, we only used the averaged spectra of selected positions in Table \ref{32s34s} to derive the $^{34}$S/$^{33}$S ratio. The averaged spectra were shown in Figure \ref{oc33s}. The $^{34}$S/$^{33}$S isotope ratio could be directly obtained from
\begin{equation}
\frac{^{34}S}{^{33}S}  \sim \frac {I(OC^{34}S) } {I(OC^{33}S) },
\end{equation}
which requires that OC$^{34}$S and OC$^{33}$S are optically thin. 

Table \ref{33s} presents the Gaussian fitting results of OC$^{33}$S. We can see from table \ref{33s} that the ratio of $^{34}$S/$^{33}$S is 6.8$\pm$1.9. This value is slightly larger than those derived from the CS and its isotoplogues toward the +50 km s$^{-1}$ Cloud and Sgr B2(N), which are 4.3$\pm$0.2 \citep{humire20} and 4.6$\pm$0.2 \citep{yu20}, respectively. It is larger than that measured toward Sgr B2(N) by \citet{yan23}, which is $\sim$2. 
 
\section{DISCUSSION}\label{sec:disc}

In this paper, we present measurements of $^{32}$S/$^{34}$S and $^{34}$S/$^{33}$S ratios toward the extended region of Sgr B2 for the first time. Observations toward Sgr B2 may suffer from optical depth of emission lines, the line forest toward Sgr B2(N), as well as the blend of absorption and emission caused by strong continuum emission, which will bring uncertainties for the measurements of sulphur isotopic ratios. The physical and chemical conditions of the extended regions is relatively simple, as they are far away from star-forming regions. We took into account the contamination of OC$^{34}$S by NH$_2$D, and then selected positions that is not contaminated by NH$_2$D emission. The optical depth of OCS are calculated by comparing with O$^{13}$CS. Four positions with optically thin OCS lines were selected for further studies. At last, a $^{32}$S/$^{34}$S ratio of 17.1$\pm$0.9 and $^{34}$S/$^{33}$S ratio of 6.8$\pm$1.9 were obtained. Our results are consistent with those obtained by \citet{humire20}. They measured the carbon and sulphur abundances toward the +50 km s$^{-1}$ Cloud and several line-of-sight clouds towards Sgr B2(N), and obtained an average $^{32}$S/$^{34}$S ratio of 17.9$\pm$5.0 toward Sgr B2(N). Thus our result confirms the termination of the decreasing tendency while approaching the Galactic centre.

\citet{goldsmith81} once reported a $^{32}$S/$^{34}$S ratio of $\sim$16 for Sgr B2 from the OCS/OC$^{34}$S ratio. This value is consistent within errors with results present here. With OCS/OC$^{34}$S, \citet{armijos15} reported $^{32}$S/$^{34}$S ratios of 8.7$\pm$1.3 in l.o.s clouds towards G+0.693. However, their observations were affected by the optical thickness of OCS, as well as band pass ripples. 

 \citet{humire20} obtained an average $^{32}$S/$^{34}$S ratio of 17.9$\pm$5.0 in the envelope of Sgr B2(N) with ALMA observations of absorption lines of CS isotopologues, which is consistent with results present here. \citet{yu20} reported a $^{32}$S/$^{34}$S ratio of 7.1$\pm$4.1 measured with IRAM 30m observations of CS isotopologues. \citet{yan23} obtained three different isotopic ratios from IRAM 30m observations of CS isotopologues toward Sgr B2(N), with isotopic ratios ranging from $\sim$10 to $\sim$26. Those results may be caused by the complex physical and kinematic environments in Sgr B2(N), in which several hot cores and UC HII regions have been detected. Abundant molecules have been detected toward Sgr B2(N) \citep{belloche13}. According to line survey of Sgr B2(N) with the IRAM 30m telescope, there are several lines near to $^{13}$CS (92494.308 MHz) and $^{13}$C$^{34}$S (90926.026 MHz), such as CH$_2$OHCHO (90922.214 MHz), CH$_3$CH$_3$CO (90929.321 MHz), CH$_3$CN, v8=2 (92491.668 MHz), and $^{13}$CH$_2$CHCN (92500.874 MHz). Both \citet{yu20} and \citet{yan23} did not take into account the line blending from nearby lines. \citet{yan23} performed independent spectral fittings for the three velocity components of Sgr B2(N). We could see from Table.A.3 in \citet{yan23} that the FWHM width of $^{13}$CS is 11.23$\pm$0.63 km s$^{-1}$ for the 53 km s$^{-1}$ component, while the FWHM of $^{13}$C$^{34}$S is 16.77$\pm$0.81 km s$^{-1}$ for the same velocity component. The FWHM of $^{13}$CS is 16.2$\pm$0.63 km s$^{-1}$ for the 82 km s$^{-1}$ component, while the FWHM of $^{13}$C$^{34}$S is 10.43$\pm$1.85 km s$^{-1}$ for the same velocity component. The significant difference in FWHM indicates that these velocity components are affected by line blending from nearby lines, thus we think that the measurements toward Sgr B2(N) are unreliable because of the serious line blending toward Sgr B2(N). Measurements toward the extended envelope away from Sgr B2(N) and Sgr B2(M) should be more reliable. 
 
With the velocity-corrected averaging spectra, we obtained a $^{34}S/^{33}S$ ratio of 6.8$\pm$ 1.9. This value is a factor of 3 higher isotopic ratio than that present in \citet{yan23}. We could see from \citet{belloche13} that there are two lines near to C$^{33}$S 2-1 (97172.064 MHz), including HCCC$^{15}$N, v=0 (97165.829 MHz) and C$_2$H$_3$CN, v15=2 (97183.100 MHz). The HCCC$^{15}$N, v=0 is only 6 MHz away from C$^{33}$S 2-1. The frequency separation is smaller than the expected FWHM width of 30 km s$^{-1}$, corresponding to a frequency range of about 10 MHz. In addition, the C$^{33}$S 2-1 line consists of eight hyperfine components distributed over about 9 MHz. The contamination from nearby lines and the hyperfine components of C$^{33}$S make it hard to determine the $^{34}S/^{33}S$ ratio with single dish observation of CS lines toward Sgr B2(N). Our OC$^{33}$S measurement is less affected by such contamination, and hence is likely to represent more accurate $^{33}$S abundance. 

The metallicity, i.e. the content of metals relative to hydrogen, could give a measure of the integrated star formation in galaxies. According to the cutting-edge model calculations \citep{kobayashi11}, the sulphur isotopic ratios are related with metallicity [Fe/H]. Based on the relationship obtained in \citet{kobayashi11, humire20}, $^{32}$S/$^{34}$ of 17.5$\pm$1.3 gives [Fe/H]  of 0.3$\pm$0.1. \citet{ryde16} obtained metallicities of $-0.2 < [Fe/H] < +0.3$ for the central 2 degrees ($\sim$ 300 pc) by observing 2 $\mu m$ spectra of M-giants with the Very Large Telescope at high spectral resolution. Thus our result is consistent with iron abundance measurement. 

\section{CONCLUSIONS}\label{sec:conclusions}

We carried out mapping observations of OCS, O$^{13}$CS, OC$^{34}$S, and OC$^{33}$S 7-6 transitions toward Sgr B2 cloud complex with the IRAM 30m radio telescope. We carefully chose positions with optically thin OCS and uncontaminated OC$^{34}$S lines, and then used their intensity ratios of the averaged spectra to derive the sulfur isotopic ratios. $^{34}$S/$^{33}$S ratio was derived using integrated intensity ratio of OC$^{34}$S and OC$^{33}$S 7-6. We obtained a $^{32}$S/$^{34}$S ratio of 17.1$\pm$0.9. This result confirms a termination of the decreasing trend of the $^{32}$S/$^{34}$S isotopic ratios when approaching the Galactic Centre. We also obtained a $^{34}$S/$^{33}$S ratio of 6.8$\pm$1.9 toward extended envelope of Sgr B2, which is consistent with that measured with CS isotopologues toward the Galactic disc \citep{chin96}.

\section*{ACKNOWLEDGEMENTS}

The authors thank the staff at IRAM for their excellent support of these observations. This work made use of the CDMS Database. This work has been supported by the National Key R\&D Program of China (No. 2022YFA1603101). The single dish data are available in the IRAM archive at https://www.iram-institute.org/EN/content-page-386-7-386-0-0-0.html. This research has made use of NASA's Astrophysics Data System.

\onecolumn

\begin{table*}
\scriptsize
      \caption{Observed Transitions}
          \begin{center}
      \label{table1}
      \begin{tabular}{lcccccc}
    \hline
    \hline
 (1) {\bf Molecule} &      (2) Transition           & (3) Rest Freq.     & (4) $E_{l}/k$   & (5) HPBW               \\
                    &                           &  (MHz)                &  (K)    &       (\arcsec)             \\
\hline        
OC$^{34}$S    &  7-6         & 83057.971(0.001)     & 11.96     &    29.6      \\     
O$^{13}$CS    &    7-6    & 84865.166(0.015)      & 12.22    &    29.0      \\     
 OCS    &   7-6        & 85139.103(0.001)    & 12.26     &    28.9      \\ 
  OC$^{33}$S    &   7-6 & 84067.082(0.001)    & 12.10     &    29.3      \\                                     
\hline
      \end{tabular}  \\
  \end{center}
  Notes.-: Lines used for mapping. Col. (1): chemical formula; Col. (2): transition quantum numbers, Col. (3): rest frequency and its uncertainty;  Col.(4): lower state energy level (K); Col. (5): Half-power beam width, which was calculated following Eq. (1) in https://www.iram.es/IRAMES/telescope/telescopeSummary/telescope\_summary.html.
\end{table*}

\clearpage

\begin{sidewaystable}
\vspace{2cm}
    \caption{Physical parameters of OCS, O$^{13}$CS, and OC$^{34}$S}
    \tabcolsep 3pt
    \centering
    \scriptsize
    \label{32s34s}
    \begin{tabular} {lcccccccccccccc}
    \hline 
    \hline
      &            &                                     &         OCS           &  7-6      &         &                    &        O$^{13}$CS     &     7-6     &        &          OC$^{34}$S    &    7-6         &                            \\
      \hline
 offset   &  $T_A^*$(OCS)  &  $V_{LSR}$   & $\Delta V$  &   $I$(OCS) & $\tau$(OCS) &  $T_A^*$(O$^{13}$CS) &  $V_{LSR}$    & $\Delta V$  & $I$(O$^{13}$CS) &  $T_A^*$(OC$^{34}$S)  &  $ V_{LSR}$  & $\Delta V$ &   $I$(OC$^{34}$S) &   R(32/34)   \\
 \arcsec  &    mK   &     km s$^{-1}$  &     km s$^{-1}$   &   K km s$^{-1}$  &  &      mK   &  km s$^{-1}$   & km s$^{-1}$  &   K km s$^{-1}$ &  mK  &     km s$^{-1}$ &  km s$^{-1}$  &  K km s$^{-1}$   &         \\
\hline
 60,-30   &   560$\pm$16   &    58.6$\pm$0.1       &    19.1$\pm$0.2  & 11.36$\pm$0.12       &  -$^a$   &   17$\pm$7  &   55.7$\pm$2.1 &  21.5$\pm$3.8  &   0.40$\pm$0.07    &   37$\pm$10   &   55.9$\pm$0.9  &  19.2$\pm$2.2    &  0.75$\pm$0.07      &    15.1$\pm$1.6                    \\
 60,-90   &  1020$\pm$12   &   61.1$\pm$0.1    &      16.9$\pm$0.1    & 18.30$\pm$0.09    &    -   &  32$\pm$9   &  59.7$\pm$1.0  &  16.9 $\pm$2.0    &  0.58$\pm$0.06   &  51$\pm$12  &  59.3$\pm$0.8 &  18.9$\pm$2.4   &   1.03$\pm$0.10    &              17.8$\pm$1.8             \\
 -90,-30   &   760$\pm$54   &   62.6$\pm$0.3      &    18.9$\pm$0.7     &  15.27$\pm$0.47    &  0.12  &    26$\pm$8  &  62.1$\pm$1.0  &  15.0 $\pm$2.0  & 0.41$\pm$0.05   & 36$\pm$10  &  62.9$\pm$1.1   &   21.5$\pm$2.8      &  0.84$\pm$0.09   &    18.2$\pm$2.5                \\
 30,-180    &  773$\pm$22   &   49.7$\pm$0.1    &      10.6$\pm$0.2    &  8.75$\pm0$.13       &   0.09      &  26$\pm$10   &   48.2$\pm$1.1  &   10.8$\pm$2.3  &   0.30$\pm$0.06  &  43$\pm$10  &  49.1$\pm$0.6   &  9.3$\pm$1.4   &  0.43$\pm$0.07  &  20.3$\pm$3.6          \\	
\hline
aver. spec.   &   736$\pm$20  &   58.8$\pm$0.1   & 16.3$\pm$0.2    &   12.8$\pm$0.14     &     -      & 23$\pm$4   &  57.6$\pm$0.8    &   17.4$\pm$1.7  &  0.43$\pm$0.04 &  39$\pm$5 &   57.4$\pm$ 0.5 &  18.1$\pm$1.3    &  0.75$\pm$0.04      &  17.1$\pm$0.9       \\ 
    \hline
    \end{tabular}      \\
 Note. Col(1): the equatorial offsets with respect to Sgr B2(N); Col(2): the peak intensity of OCS 7-6; Col(3): $V_{LSR}$ of OCS 7-6; Col(4): the FWHM of OCS 7-6; Col(5): the integrated intensity of OCS 7-6; Col(6): peak opacity of OCS 7-6; Col(7): the peak intensity of O$^{13}$CS 7-6; Col(8):  $V_{LSR}$ of O$^{13}$CS 7-6; Col(9): the FWHM of O$^{13}$CS 7-6; Col(10): the integrated intensity of O$^{13}$CS 7-6;  
 Col(11): the peak intensity of OC$^{34}$S 7-6; Col(12): the $V_{LSR}$ of OC$^{34}$S 7-6; Col(13): the FWHM of OC$^{34}$S 7-6; Col(14): the integrated intensity of OC$^{34}$S 7-6; Col(15): the isotopic ratio of $^{32}$S to $^{34}$S. $^a$: For positions where $\tau$ was unmeasurable, OCS emission was regarded to be optically thin.  
    \end{sidewaystable}
    
    \clearpage

\begin{sidewaystable}
\vspace{9cm}
    \caption{Physical parameters of OC$^{33}$S for the average spectrum}
    \centering
    \scriptsize
    \label{33s}
    \begin{tabular}{lcccccc}  
    \hline
    \hline
   &         &                    &        OC$^{33}$S                    &     7-6     &                              \\
  $T_A^*$     &  $V_{LSR}$    & $\Delta V$    &   $I$     &     R(34/33)   \\
  mK   &  km s$^{-1}$   & km s$^{-1}$  &   K km s$^{-1}$ &            \\
\hline
6.4$\pm$5.6   &  58.8   &  16.3  &  0.11$\pm$0.03  &  6.8$\pm$1.9                   \\
    \hline
    \end{tabular}      \\
 Note. Col(1): the peak intensity of OC$^{33}$S 7-6; Col(2): $V_{LSR}$ of OC$^{33}$S 7-6; Col(3): the FWHM of OC$^{33}$S 7-6; Col(4): the integrated intensity of OC$^{33}$S; Col(5): the isotopic ratio $^{34}$S to $^{33}$S.     
    \end{sidewaystable}
    
    \clearpage
    
\clearpage

\clearpage

\clearpage

\clearpage

\clearpage

\clearpage

{}

\end{document}